\documentclass[twocolumn]{aastex7}

\usepackage{amsmath}
\usepackage{textgreek}

\newcommand{\ha}{\mbox{H$\alpha$}}
\newcommand{\kms}{\ensuremath{\mathrm{km\,s^{-1}}}}
\newcommand{\mstar}{\mbox{{M${_*}$}}}
\newcommand{\msun}{\mbox{${M_\odot}$}}

\usepackage[dvipsnames]{xcolor}

\newcommand{\CIII}{C\,\textsc{iii]}}
\newcommand{\CIV}{C\,\textsc{iv}}
\newcommand{\MgII}{Mg\,\textsc{ii}\,$\lambda\lambda2796,2803$}

\newcommand{\HeII}{He\,\textsc{ii}\,$\lambda1640$}
\definecolor{forestgreen}{rgb}{0.13, 0.55, 0.13}

\begin{document}

\title{Undermassive Hosts of \mbox{\boldmath $z$\,$=$\,4--6} AGN from JWST/NIRCam Image Decomposition with \\ JADES, CONGRESS, and FRESCO
}

\author[0009-0003-5402-4809]{Zheng Ma}
\affiliation{Steward Observatory, University of Arizona, 933 North Cherry Avenue, Tucson, AZ 85721, USA}
\email[show]{mazh@arizona.edu}

\author[0000-0003-1344-9475]{Eiichi Egami}
\affiliation{Steward Observatory, University of Arizona, 933 North Cherry Avenue, Tucson, AZ 85721, USA}
\email{egami@arizona.edu}

\author[0000-0003-3307-7525]{Yongda Zhu} \thanks{JASPER Scholar}
\affiliation{Steward Observatory, University of Arizona, 933 North Cherry Avenue, Tucson, AZ 85721, USA}
\email{yongdaz@arizona.edu}

\author[0000-0002-4622-6617]{Fengwu Sun}
\affiliation{Center for Astrophysics $|$ Harvard \& Smithsonian, 60 Garden St., Cambridge, MA 02138, USA}
\email{fengwu.sun@cfa.harvard.edu}

\author[0000-0002-6221-1829]{Jianwei Lyu}
\affiliation{Steward Observatory, University of Arizona,
933 North Cherry Avenue, Tucson, AZ 85721, USA}
\email{}

\author[0000-0002-1574-2045]{Junyu Zhang}
\affiliation{Steward Observatory, University of Arizona, 933 North Cherry Avenue, Tucson, AZ 85721, USA}
\email{}

\author[0000-0001-9262-9997]{Christopher N.\ A.\ Willmer}
\affiliation{Steward Observatory, University of Arizona, 933 North Cherry Avenue, Tucson, AZ 85721, USA}
\email{}

\author[0000-0002-8651-9879]{Andrew J.\ Bunker}
\affiliation{Department of Physics, University of Oxford, Denys Wilkinson Building, Keble Road, Oxford OX1 3RH, UK}
\email{}

\author[0000-0002-6719-380X]{Stefano Carniani}
\affiliation{Scuola Normale Superiore, Piazza dei Cavalieri 7, I-56126 Pisa, Italy}
\email{stefano.carniani@sns.it}

\author[0000-0002-9551-0534]{Emma Curtis-Lake}
\affiliation{Centre for Astrophysics Research, Department of Physics, Astronomy and Mathematics, University of Hertfordshire, Hatfield AL10 9AB, UK}
\email{e.curtis-lake@herts.ac.uk}

\author[0000-0002-8543-761X]{Ryan Hausen}
\affiliation{Department of Physics and Astronomy, The Johns Hopkins University, 3400 N. Charles Street, Baltimore, MD 21218, USA}
\email{rhausen@ucsc.edu}

\author[0000-0002-1660-9502]{Xihan Ji}
\affiliation{Kavli Institute for Cosmology, University of Cambridge, Madingley Road, Cambridge, CB3 0HA, UK}
\affiliation{Cavendish Laboratory, University of Cambridge, 19 JJ Thomson Avenue, Cambridge, CB3 0HE, UK}
\email{}

\author[0000-0001-7673-2257]{Zhiyuan Ji} \thanks{JASPER Scholar}
\affiliation{Steward Observatory, University of Arizona, 933 North Cherry Avenue, Tucson, AZ 85721, USA}
\email{}

\author[0009-0003-7423-8660]{Ignas Juod\v{z}balis}
\affiliation{Kavli Institute for Cosmology, University of Cambridge, Madingley Road, Cambridge, CB3 0HA, UK Cavendish Laboratory, University of Cambridge, 19 JJ Thomson Avenue, Cambridge, CB3 0HE, UK}
\email{}

\author[0000-0002-4985-3819]{Roberto Maiolino}
\affiliation{Kavli Institute for Cosmology, University of Cambridge, Madingley Road, Cambridge, CB3 0HA, UK}
\affiliation{Cavendish Laboratory - Astrophysics Group, University of Cambridge, 19 JJ Thomson Avenue, Cambridge, CB3 0HE, UK}
\affiliation{Department of Physics and Astronomy, University College London, Gower Street, London WC1E 6BT, UK}
\email{rm665@cam.ac.uk}

\author[0000-0003-2303-6519]{George H.\ Rieke}
\affiliation{Steward Observatory, University of Arizona, 933 North Cherry Avenue, Tucson, AZ 85721, USA}
\email{grieke@arizona.edu}

\author[0000-0002-5104-8245]{Pierluigi Rinaldi}
\affiliation{Space Telescope Science Institute, 3700 San Martin Dr., Baltimore, MD 21218, USA}
\email{}

\author[0000-0001-6561-9443]{Yang Sun}
\affiliation{Steward Observatory, University of Arizona, 933 North Cherry Avenue, Tucson, AZ 85721, USA}
\email{sunyang@arizona.edu}

\author[0000-0002-8224-4505]{Sandro Tacchella}
\affiliation{Kavli Institute for Cosmology, University of Cambridge, Madingley Road, Cambridge, CB3 0HA, UK}
\affiliation{Cavendish Laboratory, University of Cambridge, 19 JJ Thomson Avenue, Cambridge, CB3 0HE, UK}
\email{st578@cam.ac.uk}

\author[0000-0003-4891-0794]{Hannah \"Ubler}
\affiliation{Max-Planck-Institut f\"ur extraterrestrische Physik (MPE), Gie{\ss}enbachstra{\ss}e 1, 85748 Garching, Germany}
\email{hannah@mpe.mpg.de}

\author[0000-0003-2919-7495]{Christina C.\ Williams}
\affiliation{NSF National Optical-Infrared Astronomy Research Laboratory, 950 North Cherry Avenue, Tucson, AZ 85719, USA}
\email{christina.williams@noirlab.edu}

\begin{abstract}
    In the local Universe, supermassive black hole (SMBH) masses strongly correlate with their host-galaxies' stellar masses ($M_{*}$), but galaxies hosting faint AGN recently found by JWST may deviate from this relation. To constrain the M$_{\text{BH}}$-M$_{*}$ relation at high redshift, we performed AGN-host image decomposition for 17 low-luminosity AGN galaxies at $z$\,$\sim$\,4--6 using NIRCam images in the JADES GOODS-N field. These sources are identified as AGN from broad H$\alpha$ emission lines detected by the CONGRESS and FRESCO surveys. We used \textsc{galfit+MCMC} to fit spatial profiles in 4--7 wide-band images and detected extended emission in 9 sources out of 17. The close spatial alignment between the extended S\'ersic-profile components and the centers of poine-like AGN sources hints that this emission likely originates from the host galaxies and not from AGN-excited nebular emission. These sources are extended at 0.9--2.0~$\mu$m, suggesting significant host-galaxy light in the rest-frame UV. For the sources with the host detection, the stellar mass inferred from the image decomposition analysis can be 1-2 dex lower than the results without image decomposition. The BH-to-stellar mass ratio spans $M_{\text{BH}}/M_\ast$\,$\sim$\,0.01--1.48, placing them well above the local $M_{\text{BH}}$--$M_\ast$ relation. In contrast, the host-galaxy size--mass relation broadly agrees with previous measurements, supporting our host-galaxy interpretation. Our results suggest that the host galaxies of these faint AGN are either genuinely under-massive compared to their black hole masses, or too compact to be spatially resolved with the NIRCam imaging data.
\end{abstract}

\section{Introduction}

In the local Universe, a well-established correlation exists between the masses of supermassive black holes (SMBHs) and host galaxy properties, such as bulge mass and velocity dispersion \citep{Magorrian_1998, Ferrarese_2000,Gebhardt_2000,Kormendy_2013}. These relationships are widely interpreted as evidence for the co-evolution of SMBHs and their host galaxies. Among these relations, one particularly interesting connection is the stellar mass -- black hole mass ($M_*$--$M_{\rm BH}$) relation, which implies that the growth of SMBHs and that of their host galaxies are intertwined. There are several theoretical models attempting to explain this correlation. For example, AGN can regulate host-galaxy growth by heating or expelling its surrounding gas \citep{somerville_2008,Fabian, Fan_2023}. Other hypotheses suggest that the statistical effects of major and minor galaxy mergers naturally lead to a tight, linear correlation between SMBH masses and galaxy bulge masses, regardless of how the black holes were originally seeded \citep{peng_2007}, although simulations show that for massive galaxies (\mstar $> 10^{10.9} \msun $), mergers are not likely the primary drivers of AGN activity \citep{ Sharma_2024}.

To understand the origin of the $M_*$--$M_{\rm BH}$ relation, it is necessary to examine this relation for black holes in their early stages. To derive this relation at high redshift, some studies have focused on AGN-host galaxy decompositions, particularly through image decomposition techniques \citep[e.g.,][]{Ding_2023, Yue_2024, harikane_23, Chen_2024, yu_decomposition_2025, Stone_2023}, fitting AGN-galaxy images with a combination of a single S\'ersic profile and that of a point-source (i.e., a point spread function). However, observing this relation becomes increasingly challenging at high redshift, where galaxies hosting AGN are typically more compact. Their stellar light is often outshined by the AGN, making it difficult to separate the host galaxy from the AGN component. 

The James Webb Space Telescope (JWST) presents a unique opportunity to overcome these challenges with its high spatial resolution and great sensitivity in the near-infrared. In recent years, JWST has demonstrated its capability to identify UV-faint AGN-host galaxies at high redshift ($z$\,$>$\,5) through the detection of broad Balmer emission lines \citep{matthee, maiolino_24, Ignas_2025, greene_24, ubler_2023, larson_2023, kocevski_2023, harikane_23, kokorev_2024}. These UV-faint AGN appear to be significantly more abundant than previously expected---their number density at $z$\,$\sim$\,5 is measured to be one to two orders of magnitude higher than the estimates based on the extrapolations of the quasar UV luminosity function \citep{matthee, maiolino_24, Ignas_2025, harikane_23, kocevski_2023, greene_24}. These low-luminosity, broad-line AGN galaxies typically host moderate-mass black holes with masses in the range of $10^{6-8}$ \msun, derived from the locally calibrated single-epoch methods or direct black-hole mass measurement \citep{matthee, maiolino_24, harikane_23, Ignas_2025_bhmass, Geris_2026}. In comparison, luminous quasars at similar redshifts are found to host black holes with masses of $\sim 10^{8-10}M_\odot$ \citep{Fan_2023, trakhtenbrot_11}. UV-faint AGN therefore provide ideal laboratories for constraining host-galaxy properties and investigating the co-evolution of supermassive black holes and their hosts. First, their relatively low black hole masses compared to luminous quasars suggest that they represent an earlier stage of black hole growth. Second, their low AGN luminosities reduce the AGN-host contrast, making it easier to detect and characterize the underlying host galaxies without being completely outshined by the AGN. 

About 15 - 30\% of broad-line AGN identified by JWST are ``little red dots'' (LRDs; \citealp{matthee, Kevin_2025}). Currently, there is no unified set of selection criteria for identifying these objects. Spectroscopically, they exhibit a distinct ``V-shaped'' continuum, with a red slope in the rest-frame optical and a blue slope in the rest-frame UV \citep{Furtak_2024}. Morphologically, they are characterized by their compact point-like sources in the rest-frame optical with red colors \citep{kokorev_2024, Rinaldi_2025}. 

The nature of LRDs remains under debate. They have been interpreted as dust-reddened AGN based on the presence of broad Balmer emission lines and their red rest-frame optical colors \citep{matthee,kocevski_2023}. Alternatively, the presence of Balmer break and some SED-fitting studies suggest that LRDs may instead be extremely compact galaxies undergoing intense starburst activity, but this would lead to very high stellar mass estimates ($\sim 10^{10} \msun\ $) that are incompatible with the $\Lambda$CDM cosmological model \citep{Labbe_2023_nat, Akins_2023, Pablo_2024, Barro_2024, Baggen_2024}. More recently, the strong Balmer absorption feature seen in some of LRDs is taken to suggest that they are black holes embedded in dense compton-thick, dust-free gas \citep{Inayoshi_2025,Naidu_2025,deGraaff_2025,Rusakov_2025,Maiolino_gas, Rusakov_2025, Ji_2025, Lin_2025}, which may also explain their observed X-ray weakness due to heavy gas absorption. At present, there is no consensus on the physical nature of LRDs, and it remains unclear whether they represent an early phase of black hole growth or arise from other physical processes.

In this paper, we aim to decompose the host galaxy from the AGN component spatially and investigate the properties of the host galaxies, focusing particularly on their stellar masses and the relationship with the black hole masses. This will allow us to examine whether the $M_*$--$M_{\rm BH}$ \ relation co-evolves across cosmic time. In Section \ref{sec:data}, we introduce our dataset and our sample. In Section \ref{sec:method}, we describe the methods used for the image decomposition. In Section \ref{sec:results} we present the decomposition results. We discuss the implications in Section \ref{sec:discussion} and summarize our findings in Section \ref{sec:conclusion}. Uncertainties associated with the image-decomposition analysis and the robustness of the host-flux recovery are presented in Appendix~\ref{flux-recovery}.

Throughout this work, we use the AB magnitude system \citep{oke_gunn83}
and assume a flat $\Lambda$CDM cosmology with $\Omega_m$ = 0.3 and $H_0 = 70 $ km s$^{-1}$ Mpc$^{-1}$. 

\section{Data and Sample Selection}
\label{sec:data}

The Imaging data used in this work are obtained by the JWST Advanced Deep Extragalactic Survey (JADES, \citealp{Bunker_2020,DEugenio_2025,CurtisLake_2025, Scholtz_2025,fitsmap,Johnson2026,Robertson2026,Carreira2026}). Data used in this work are released as part of the JADES DR5 data release \citep{Robertson2026,Johnson2026}. We utilize NIRCam imaging in seven wide-band filters (F090W, F115W, F150W, F200W, F277W, F356W, F444W). The native pixel scale of the NIRCam short-wavelength (SW) channel is 0\farcs031 per pixel, while the long-wavelength (LW) channel has a native scale of 0\farcs063 per pixel. All science mosaics analyzed here have been drizzled to a common pixel scale of 0\farcs03 per pixel to facilitate multi-band alignment and consistent photometric and morphological measurements.
 
\cite{Zhang_2025} searched for broad \ha\ emitters in the GOODS-N field using JWST/NIRCam slitless spectroscopy from FRESCO (``First Reionization Epoch Spectroscopically Complete Observations'', PID: 1895, PI: P. Oesch, \citealp{Oesch_2023}) and CONGRESS (``Complete NIRCam Grism
Redshift Survey'', PID: 3577, PI: E. Egami; F. Sun et al., in prep). \cite{Zhang_2025} identified 19 broad \ha\ emitters at z\,$\sim$\,4--5.5 with FWHM $>$ 1000 \kms\ and interpreted them as Type-I AGN galaxies; they also measured the black-hole masses of these AGN galaxies based on the luminosity and FWHM of the broad \ha\ component. All of the 19 AGN galaxies are powered by relatively low-mass black holes with masses ranging from $10^{6.65}$ to $10^{8.29}$ \msun, which are about two orders of magnitude lower than those of typical quasars \citep{Fan_2023} but comparable to faint AGN reported by recent JWST surveys \citep{matthee, greene_24, maiolino_24}. Although the JADES dataset covers both the GOODS-N and GOODS-S fields, we restrict this study to the GOODS-N field because of the wider wavelength coverage afforded by the combination of the CONGRESS and FRESCO data. Identification of additional AGN galaxies in the GOODS-S field is being conducted with the new GTO NIRCam Grism F356W spectroscopy (PI: G.~Rieke; PID-4549). 

All 19 broad Type-I AGN galaxies lie within the JADES footprint. Among them, GN1034620 overlaps with the edge of a foreground galaxy, and GN9994014 is contaminated by some image artifacts. We therefore exclude these two sources from our analysis. For the remaining 17 galaxies, the image quality is sufficient for reliable image decomposition.  Table~\ref{table:decomposition_result} lists the final sample of 17 Type-I AGN galaxies in GOODS-N analyzed in this paper.  

\begin{figure*}
    \centering
    \includegraphics[width=0.7\linewidth]{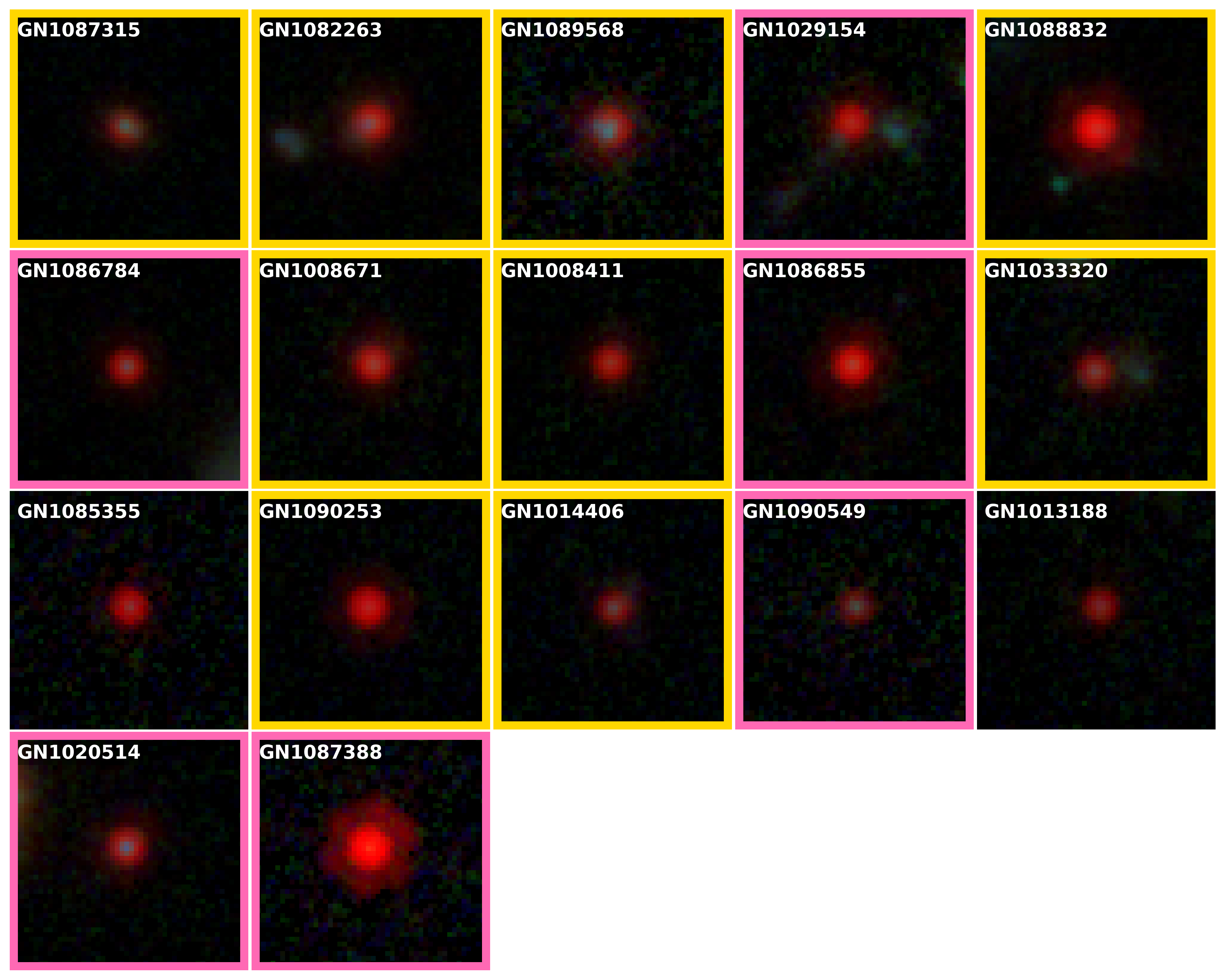}
    \caption{False-color postage-stamp images ($1\farcs5\,\times\,1\farcs5$) of the 17 Type-1 AGN galaxies produced with the JADES NIRCam data. The RGB images are constructed by combining F090W+F115W (blue), F150W+F200W (green), and F277W+F356W+F444W (red). For GN1089568, GN1085355, GN1090549, and GN1087388, F090W, F115W, and the average of F356W and F444W are used for the blue, green, and red channels, respectively. Nine sources with strong evidence of host detection are marked in gold while six sources with tentative host detection are marked in pink; Two sources with no color border (GN1085355 and GN1013188) do not show any sign of host detection.}
    \label{fig:rgbcollage}
\end{figure*}

Among the sample, GN1089668, GN1087388, and GN1090549 have NIRCam images available in only four bands (F090W, F115W, F356W, and F444W). GN1090253 has five bands available (F090W, F115W, F150W, F356W, and F444W), while GN1086855 has six bands, missing only F277W.  For all other sources, imaging data are available in all seven NIRCam bands. We present the false-color stamp images of the 17 Type-1 AGN galaxies in Figure~\ref{fig:rgbcollage}. 

\begingroup
\setlength{\tabcolsep}{7pt}
\begin{deluxetable*}{lccccccccc}
\caption{Decomposition results of Type-I (broad-line) AGN galaxies in the JADES GOODS-N field}
\label{table:decomposition_result}
\tablehead{
  \colhead{JADES ID} &
  \colhead{$z_{\rm spec}$} &
  \colhead{$\Delta$BIC} &
  \colhead{Flag} &
  \colhead{Filter} &
  \multicolumn{2}{c}{$R_e$} &
  \colhead{$n$} &
  \multicolumn{2}{c}{Center Offset} \\
  \cline{6-7}
  \cline{9-10}
  \colhead{} &
  \colhead{} &
  \colhead{} &
  \colhead{} &
  \colhead{} &
  \colhead{(\arcsec)} &
  \colhead{(kpc)} &
  \colhead{} &
  \colhead{(\arcsec)} &
  \colhead{(kpc)}
}
\decimalcolnumbers
\startdata
GN1087315 & 3.91 & -9662 & Y & F150W & 0.063$\pm$0.003 & 0.44$\pm$0.02 & 1.20$\pm$0.19 & 0.030 & 0.21 \\
GN1082263 & 3.98 & -16189 & Y & F150W & 0.058$\pm$0.004 & 0.40$\pm$0.03 & 2.60$\pm$0.38 & 0.034 & 0.24 \\
GN1089568 & 4.05 & -6471 & Y & F090W & 0.045$\pm$0.001 & 0.31$\pm$0.01 & 0.81$\pm$0.08 & 0.032 & 0.22 \\
GN1029154 & 4.17 & -257 & X & F115W & 0.124$\pm$0.230 & 0.85$\pm$1.57 & 4.90$\pm$6.60 & 0.016 & 0.11 \\
GN1088832 & 4.36 & -11240 & Y & F115W & 0.101$\pm$0.017 & 0.68$\pm$0.11 & 4.35$\pm$0.68 & 0.030 & 0.20 \\
GN1086784 & 4.40 & -186 & X & F200W & 0.060$\pm$0.038 & 0.40$\pm$0.26 & 8.00$\pm$7.15 & 0.009 & 0.06 \\
GN1008411 & 4.41 & -1719 & Y & F115W & 0.089$\pm$0.008 & 0.60$\pm$0.05 & 0.70$\pm$0.17 & 0.060 & 0.40 \\
GN1086855 & 4.41 & -177 & X & F150W & 0.240$\pm$0.733 & 1.60$\pm$4.89 & 8.00$\pm$13.90 & 0.018 & 0.12 \\
GN1008671 & 4.41 & -1681 & Y & F150W & 0.037$\pm$0.008 & 0.24$\pm$0.06 & 4.60$\pm$1.87 & 0.007 & 0.04 \\
GN1033320 & 4.48 & -750 & Y & F115W & 0.131$\pm$0.031 & 0.87$\pm$0.20 & 1.73$\pm$0.48 & 0.017 & 0.11 \\
GN1085355 & 4.88 & 20 & N & F115W & - & - & - & - & - \\
GN1090253 & 5.09 & -1091 & Y & F150W & 0.033$\pm$0.022 & 0.21$\pm$0.14 & 4.27$\pm$4.65 & 0.004 & 0.03 \\
GN1014406 & 5.15 & -1031 & Y & F115W & 0.123$\pm$0.015 & 0.76$\pm$0.09 & 0.70$\pm$0.22 & 0.055 & 0.34 \\
GN1090549 & 5.20 & -87 & X & F090W & 0.057$\pm$0.015 & 0.35$\pm$0.09 & 0.70$\pm$0.88 & 0.002 & 0.01 \\
GN1013188 & 5.25 & 34 & N & F150W & - & - & - & - & - \\
GN1020514 & 5.36 & -431 & X & F115W & 0.016$\pm$0.003 & 0.10$\pm$0.02 & 1.08$\pm$0.70 & 0.006 & 0.04 \\
GN1087388 & 5.54 & -487 & X & F090W & 0.111$\pm$0.050 & 0.66$\pm$0.30 & 1.79$\pm$1.13 & 0.021 & 0.12 \\
\enddata
\tablecomments{
Columns:
(1) IDs from the JADES DR2 \citep{DEugenio_2025}.
(2) Spectroscopic redshifts from \cite{Zhang_2025}.
(3) BIC difference between the fits using the S\'ersic+PS model and PS-only model;
$\Delta\mathrm{BIC} \equiv \mathrm{BIC}_{\rm (PS+Sersic)} - \mathrm{BIC}_{\rm (PS)}$.
(4) Flag indicating the significance of the host-galaxy detection based on $\Delta\mathrm{BIC}$:
``Y'' for sources with confident detectable extended emission ($\Delta\mathrm{BIC} < -500$),
``X'' for sources with tentative evidence of host detection ($-500 < \Delta\mathrm{BIC} < 0$),  
and ``N'' for sources with no host detection ($\Delta\mathrm{BIC} > 0$).
(5) Filter of the imaging band used to determine the S\'ersic profile.
(6) Effective radius $R_e$ in arcsec with 1 $\sigma$ statistical uncertainties from \textsc{GALFIT}, which do not account for some systematic effects such as insufficient pixel sampling of the surface brightness profiles.
(7) Same as (6) but in kpc.
(8) S\'ersic index $n$ with 1 $\sigma$ uncertainties.
(9) Offsets between the centers of the PS and S\'ersic components in arcsec.
(10) Same as (9) but in kpc.
}
\end{deluxetable*}
\endgroup

\section{AGN-Host image decomposition}
\label{sec:method}
We first perform AGN--host image decomposition using \textsc{galfit} \citep{peng2002, peng2010}. For each source, we extract a 50 × 50 pixel cutout (1.5$^{\prime\prime}$ × 1.5$^{\prime\prime}$) centered on the source. Although high-redshift galaxies are often characterized by irregular spatial distributions of stellar mass and light \citep{Huertas_2024} and may exhibit multiple star-forming clumps \citep{Fujimoto_2024}, our sources appear compact and show no obvious substructures in their images. They are only marginally resolved in the short-wavelength (SW) bands. We therefore model the AGN component using a point spread function (PSF) for each band and represent the host galaxy with a single S\'ersic profile convolved with the same PSF, a common approach among AGN--host decomposition studies \citep{Ding_2023, harikane_23, zhuang_2024, Chen_2024}. Our primary goal is not to resolve the internal structures of the host galaxies but rather to obtain an overall characterization of their global properties, particularly their stellar masses.

The PSF models used are mosaic PSFs generated by the JADES Team \citep{Robertson2026}, derived by injecting model PSFs calculated using STPSF \citep{Perrin2014} into individual exposures, which are subsequently drizzled together through the mosaicking procedure. These models account for both observational effects (e.g., source position on the detector) and data reduction processes (e.g., mosaicking).

We first fit the F444W cutout image using only a PSF model and use the resulting center coordinates as the AGN center across all bands. Then we fit each band separately with \textsc{galfit} using a PSF model placed at the fixed center and a S\'ersic model whose center we allow to vary.
During the fitting, the S\'ersic index $n$ is restricted to a range of 0.7--8, and the effective radius $R_e$ is constrained to 0.1--8 pixels (0\farcs003--0\farcs24, corresponding to $\sim$\,0.02--1.46 kpc at $z$\,$=$\,5). We allow the axis ratio to vary between 0.1--1. These parameter ranges are consistent with those used by 
previous studies \citep[e.g.,][]{zhuang_2024}.

We selected the best-fit S\'ersic model in one of the SW bands to serve as the representative model for the host galaxy through all NIRCam bands. This selection was based on both visual inspection and comparing the reduced $\chi^2$. This \textsc{galfit}-derived best-fit model was adopted and convolved with the PSFs to generate a S\'ersic model in each band.

We then use these S\'ersic models, along with the point source models (PSFs), as inputs to run MCMC (Markov Chain Monte Carlo) fitting across all bands independently to determine the flux contribution from each component. We use the affine-invariant ensemble MCMC sampler implemented in emcee, keeping the PSF-based AGN model and the PSF-convolved S\'ersic model fixed and allowing the MCMC to determine only their scaling coefficients and a constant background term. This yields the amount of light contributed by the point source and the S\'ersic component in each band. AGN (i.e., point-source) fluxes are derived accordingly, and AGN contributions are subtracted from the original images to isolate extended emission.

Fluxes of the extended emission are measured via aperture photometry on the AGN-subtracted images, while total fluxes are measured on the original images using the same aperture sizes. We use a circular aperture with a radius of 0\farcs45 and a sky annulus with inner and outer radii of 0\farcs60and 0\farcs75, respectively. For sources that are blended with nearby contaminants, they are modeled with additional S\'ersic components using the same constraints described above. In such cases, aperture photometry would include contaminating flux from nearby sources; Therefore, we take the total flux of the S\'ersic component associated with the target galaxy as the extended-emission flux. An aperture correction was calculated for each band based on its corresponding PSF. For the values of aperture correction, see J. Trussler et al.\ (in preparation).

Note that the S\'ersic model used to describe the extended emission during MCMC fitting is assumed to have the same S\'ersic parameters across all bands. The S\'ersic index and effective radius of galaxies are known to vary with wavelength \citep{Vulcani_2014,kennedy_2015}, although a recent study shows no significant differences between the stellar continuum sizes at different wavelengths from UV to optical \citep{Lola_2025}. Some fitting packages, such as \textsc{GALFITM} \citep{galfitm_boris, galfitm_vika} and \textsc{GALIGHT} \citep{galight}, attempt to account for the wavelength dependence of S\'ersic parameters by performing simultaneous multi-band fitting and linking these parameters across bands using polynomials. However, these approaches also have limitations.  More specifically, the S\'ersic parameters are hard to recover at LW because,  (1) FWHM is larger at LW, severely limiting the spatial resolution, and  (2) the AGN dominates at LW and the resultant high AGN/host flux ratio makes the estimates of S\'ersic parameters deviate from the true values, as we demonstratein Appendix~\ref{flux-recovery}. 

When applying \textsc{GALFITM} following the setup described in \cite{zhuang_2024}, we find that for some sources the fits in the F356W and F444W bands converge to unphysical solutions, characterized by extremely small effective radii ($\sim$ 10\% of the PSF FWHM) and very large S\'ersic indices that reach the imposed parameter boundaries. This behavior is driven by the limited constraining power of the data on the host-galaxy structural parameters in the LW bands. In such cases, the inferred flux contribution of the Sérsic component at F444W can increase by more than 100\% compared to the results obtained when fixing the Sérsic parameters from an SW band. For sources where GALFITM does not break down at long wavelengths—i.e., where it yields physically reasonable Sérsic profiles across all bands — we find consistent results between the two approaches. For example, in the case of GN1014406, GALFITM yields a stellar mass estimate of $\log(\mstar) = 7.68^{+0.12}_{-0.09}$, which is consistent with our result of $\log(\mstar) = 7.64^{+0.20}_{-0.23}$ obtained using the GALFIT + MCMC method. Therefore we first use GALFIT to extract the structural information of the extended emission from an SW band, taking advantage of its higher spatial resolution, narrower PSF, and lower AGN contribution compared to the LW bands, and then fix this Sérsic profile across all bands and use MCMC fitting to decompose the flux contributions of the point source and extended components. 

We compared our method with \textsc{Forcepho} \citep{forcepho}, which was used by \cite{Ignas_2025} to decompose faint AGN-hosting galaxies and derive stellar masses. Although there is no overlap between our sample and theirs, we applied our method to the most extended source in their sample, GN53757. We obtained a stellar mass of $\log(M_\ast/M_{\odot}) = 8.83^{+0.19}_{-0.10}$. For the same object, \cite{Ignas_2025} reported $\log(M\ast/M_{\odot}) = 8.81^{+0.26}_{-0.26}$ based on \textsc{Forcepho}. The close agreement between these independent measurements indicates that different image-fitting approaches produce consistent stellar mass estimates, supporting the robustness of our method.

To determine whether extended emission is detected for each target, we also perform image fitting using a single point-source (PS) model. We then evaluate whether adding the additional S\'ersic component significantly improves the fit compared to using the PS model alone. The relative goodness of each fit is quantified using the Bayesian Information Criterion (BIC)
\begin{equation}
    BIC = \chi^2 + k\ln{n},
\end{equation}
where $k$ is the number of free model parameters, and $n$ is the total number of data points used in the fit \citep{Schwarz_1978}. For this work, our fitting has 3 free parameters (flux of S\'ersic model, AGN model and background) when applying a composite S\'ersic + PS model and 2 free parameters (flux of AGN model and background) when a PS-only model is applied.
$\chi^2$ measures the residual error of the model. Lower BIC values indicate a preferred balance between fit quality and model simplicity.

\section{Results}
\label{sec:results}
\subsection{AGN-Host Image Decomposition}

\begin{figure*}[!ht]
    \centering
    \includegraphics[width=0.79\linewidth]{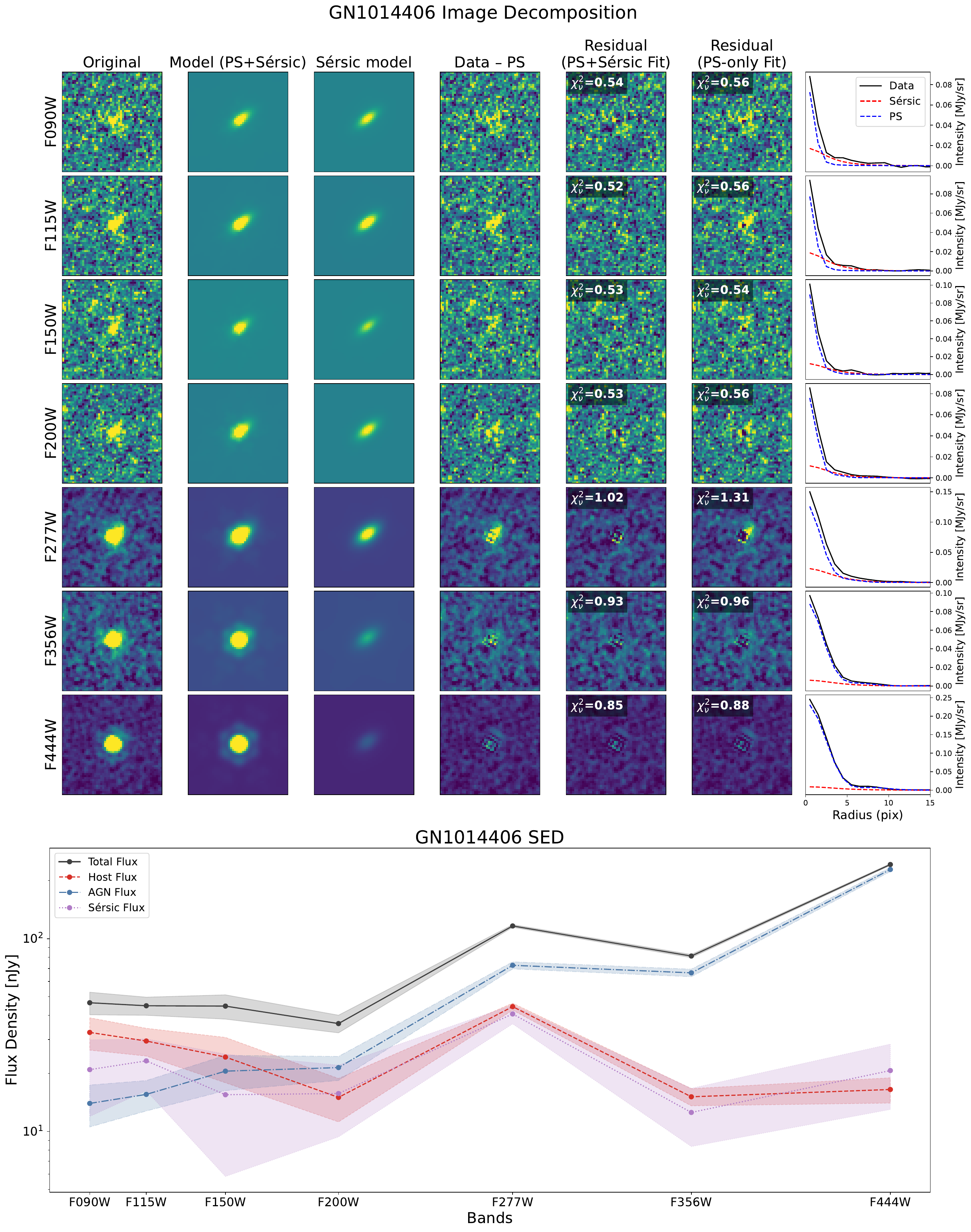}
    \caption{\textbf{Top panel:} AGN-Host decomposition of GN1014406. The cutout image size is $1.5^{\prime\prime}\times 1.5^{\prime\prime}$. The first column displays the original image in each band. The second column shows the two component model (PSF + S\'ersic). The third column shows the S\'ersic (i.e., host-galaxy) model. The fourth column shows the AGN-subtracted images, which show some residual of extended emission in the SW bands as a signal of host detection.  The fifth column shows the residual images after subtracting the PSF$+$S\'ersic model, with the fitting reduced $\chi^2$ denoted in each image. The sixth column shows the residual images when we attempt to fit the source with one single PSF model, with the fitting reduced $\chi^2$ denoted. The last column displays the radial brightness profile of the original data (the black solid line), S\'ersic model (the red dashed line), and PSF AGN model (the blue dashed line). 
    \textbf{Bottom panel:} SED of GN1014406. The black solid line shows the total flux, and the blue dash-dotted line shows the flux of the point-source AGN model.  The red dashed line shows the host flux measured from the AGN-subtracted images using aperture photometry while the purple dotted line shows the flux of the S\'ersic model, respectively. The shaded bands indicate the $1\sigma$ uncertainties. The full set of decomposition results for all sources is available as a figure set.\footnote{The figures are also available \href{https://github.com/plusante/Decomposition-result}{here}}
    }
    \label{fig:GN1014406_2D_decomp}
\end{figure*}

Figure~\ref{fig:GN1014406_2D_decomp} presents the fitting results for GN1014406 as an example. We observe extended emission in the AGN-subtracted images in the SW bands, where we believe that the host galaxy has not yet been outshined by the AGN.  This suggests that the emission cannot be fully explained by a point source alone. The PS+S\'ersic model also gives a significantly lower BIC than the single PS model ($\Delta\mathrm{BIC}$\,$\equiv$\,$\mathrm{BIC}_{\rm (PS+Sersic)} - \mathrm{BIC}_{\rm (PS)}$\,$=$\,$-1031$) , providing strong statistical evidence for the presence of extended emission in GN1014406.

In Table \ref{table:decomposition_result}, we list the BIC difference between the PS-only model and the composite PS + S\'ersic model for each source.  We also report the S\'ersic index and half-light radius of the best-fit S\'ersic model we adopted for each source. 

Among the 17 sources, 15 show lower BIC values when fitted with a composite PS + S\'ersic model compared to a PS-only model, indicating that the composite model provides a better description of their surface brightness profiles and that extended emission is detected. 9 of these have $\Delta\mathrm{BIC}$\,$<$\,$-500$, which we interpret as strong detections of extended emission; this is also visually evident in the residual images of the PS-only fits. There are 6 targets that have $-500$\,$<$\,$\Delta\mathrm{BIC}$\,$<$\,$0$, where the BIC favors the composite model, but the residual images do not provide clear visual confirmation of extended emission. For the remaining two targets, the PS-only model yields a lower BIC (i.e., $\Delta\mathrm{BIC}$\,$>$\,0), implying that their light distributions can be better described by a single point source.

When fitting our targets with a PSF and a S\'ersic component, we did not enforce a common centroid for the two models. Nevertheless, we find that the spatial offset between the PSF and S\'ersic centers is small, with a typical offset of $<$\,0\farcs06. Given this small offset, the extended emission is more likely associated with the host galaxy than with spatially offset nebular emission, as suggested for some LRDs \citep[e.g.,][]{Chen_2025}.

A common concern in image fitting is PSF mismatching. We assessed this effect by repeating the decomposition with an independent PSF constructed for GOODS-S in \cite{Ji_2024}, rotated to match the GOODS-N position angle. Because this PSF was derived from a different field and stacking procedure, it represents a conservative test of systematic uncertainty. The resulting stellar masses differ by at most $\lesssim 0.3$ dex, comparable to typical SED-fitting uncertainties, and therefore do not change our main conclusions.

In Table~\ref{table:host_agn_photometry}, we list the measured fluxes of the point-source and extended-emission components, together with their $1\sigma$ uncertainties. For GN1087388 and GN1087388, nearby sources blend with the target galaxy. Spectroscopic observations from CONGRESS and FRESCO confirm that these objects are foreground interlopers. To avoid contamination in the extended-emission fluxes, we take the S\'ersic-model flux associated with the target galaxy instead of aperture-photometry measurements. For GN1087388, we adopt the S\'ersic model flux for the extended emission because the PS signal is over-subtracted in some images, which would produce negative aperture-photometry measurements.

\begin{deluxetable*}{lcccccccc}
\tablewidth{0pt}
\caption{Photometry from Imaging decomposition.}
\label{table:host_agn_photometry}
\tablehead{
  \colhead{JADES ID} & \colhead{Component} & \colhead{F090W} & \colhead{F115W} & \colhead{F150W} & \colhead{F200W} & \colhead{F277W} & \colhead{F356W} & \colhead{F444W}\\
  \colhead{} &\colhead{}& \colhead{[nJy]} & \colhead{[nJy]} &\colhead{[nJy]}& \colhead{[nJy]} & \colhead{[nJy]} & \colhead{[nJy]} & \colhead{[nJy]}
}
\startdata
GN1087315 & Extended Emission & 45.1$\pm$3.6 & 35.5$\pm$3.2 & 43.2$\pm$3.2 & 53.6$\pm$3.7 & 64.0$\pm$0.9 & 53.9$\pm$1.0 & 25.9$\pm$1.1 \\
 & Point Source & 18.3$\pm$2.7 & 18.5$\pm$2.4 & 22.2$\pm$2.7 & 29.7$\pm$4.0 & 83.4$\pm$2.3 & 163.6$\pm$3.1 & 107.0$\pm$3.7 \\
\hline
GN1082263 & Extended Emission & 76.0$\pm$2.5 & 76.7$\pm$3.2 & 69.5$\pm$4.1 & 90.1$\pm$2.6 & 112.1$\pm$1.5 & 88.4$\pm$1.2 & 56.0$\pm$1.4 \\
 & Point Source & 17.6$\pm$2.0 & 18.9$\pm$2.4 & 21.2$\pm$3.5 & 44.5$\pm$3.1 & 205.6$\pm$4.5 & 377.5$\pm$4.4 & 422.1$\pm$5.1 \\
\hline
GN1089568 & Extended Emission & 170.6$\pm$7.2 & 181.7$\pm$6.0 & - & - & - & 142.2$\pm$2.9 & 82.8$\pm$5.9 \\
 & Point Source & 35.3$\pm$6.3 & 39.3$\pm$5.3 & - & - & - & 356.3$\pm$11.0 & 314.9$\pm$22.2 \\
\hline
GN1029154 & Extended Emission & 13.8$\pm$3.3 & 11.1$\pm$2.6 & 6.4$\pm$4.0 & 8.3$\pm$4.7 & 32.4$\pm$3.9 & 45.6$\pm$3.7 & 42.0$\pm$6.2 \\
 & Point Source & 3.3$\pm$1.5 & 4.2$\pm$1.2 & 9.1$\pm$2.1 & 19.6$\pm$2.6 & 111.7$\pm$2.4 & 333.1$\pm$2.5 & 394.8$\pm$4.1 \\
\hline
GN1088832 & Extended Emission & 43.0$\pm$3.6 & 46.8$\pm$2.1 & 45.3$\pm$3.0 & 52.1$\pm$3.9 & - & 105.7$\pm$0.9 & 102.7$\pm$1.2 \\
 & Point Source & 4.0$\pm$2.6 & 11.2$\pm$1.6 & 15.3$\pm$2.5 & 22.8$\pm$4.4 & - & 250.0$\pm$3.1 & 1161.5$\pm$5.0 \\
\hline
GN1086784 & Extended Emission & 7.3$\pm$2.3 & 8.9$\pm$3.3 & 4.5$\pm$3.9 & 10.7$\pm$2.6 & 18.9$\pm$1.2 & 13.3$\pm$1.4 & 18.1$\pm$1.3 \\
 & Point Source & 15.3$\pm$2.9 & 19.6$\pm$4.3 & 20.2$\pm$6.1 & 25.0$\pm$5.1 & 107.7$\pm$5.3 & 281.4$\pm$7.7 & 185.7$\pm$7.8 \\
\hline
GN1086855 & Extended Emission & 24.2$\pm$2.7 & 0.6$\pm$4.0 & 21.1$\pm$6.1 & 10.0$\pm$3.0 & - & 64.9$\pm$2.8 & 5.7$\pm$2.1 \\
 & Point Source & 19.5$\pm$2.4 & 21.3$\pm$3.7 & 28.6$\pm$6.0 & 45.5$\pm$3.8 & - & 515.3$\pm$10.0 & 439.0$\pm$4.0 \\
\hline
GN1008671 & Extended Emission & 42.7$\pm$4.4 & 50.8$\pm$3.3 & 52.4$\pm$5.5 & 76.6$\pm$4.6 & 117.9$\pm$1.9 & 94.2$\pm$1.6 & 118.8$\pm$2.5 \\
 & Point Source & 0.6$\pm$2.3 & 0.5$\pm$1.8 & 2.5$\pm$6.4 & 5.0$\pm$9.3 & 115.1$\pm$11.5 & 365.1$\pm$12.4 & 291.3$\pm$21.3 \\
\hline
GN1008411 & Extended Emission & 8.6$\pm$5.2 & 6.6$\pm$2.3 & 18.8$\pm$3.4 & 16.3$\pm$6.3 & 32.8$\pm$1.9 & 26.3$\pm$0.9 & 30.6$\pm$1.0 \\
 & Point Source & 6.1$\pm$3.3 & 7.6$\pm$1.3 & 11.1$\pm$2.1 & 20.3$\pm$5.3 & 95.7$\pm$3.4 & 231.0$\pm$2.1 & 270.0$\pm$2.2 \\
\hline
GN1033320 & Extended Emission & 23.0$\pm$4.1 & 32.4$\pm$2.9 & 23.3$\pm$4.2 & 29.5$\pm$2.9 & 42.0$\pm$2.6 & 38.5$\pm$2.6 & 16.9$\pm$4.0 \\
 & Point Source & 14.2$\pm$1.4 & 12.5$\pm$1.0 & 16.5$\pm$1.6 & 23.6$\pm$1.2 & 99.3$\pm$1.6 & 194.1$\pm$1.7 & 168.5$\pm$2.6 \\
\hline
GN1090253 & Extended Emission & 38.2$\pm$7.2 & 18.3$\pm$2.3 & 12.5$\pm$3.6 & - & - & 83.4$\pm$1.1 & 62.6$\pm$1.2 \\
 & Point Source & 1.9$\pm$5.6 & 4.7$\pm$3.6 & 5.5$\pm$6.4 & - & - & 202.7$\pm$8.2 & 432.2$\pm$10.1 \\
\hline
GN1014406 & Extended Emission & 32.6$\pm$6.2 & 29.5$\pm$4.8 & 24.3$\pm$6.4 & 15.0$\pm$3.8 & 44.4$\pm$1.9 & 15.1$\pm$1.6 & 16.5$\pm$2.5 \\
 & Point Source & 14.0$\pm$3.4 & 15.6$\pm$2.8 & 20.6$\pm$4.3 & 21.5$\pm$3.1 & 72.8$\pm$3.1 & 66.5$\pm$3.0 & 228.5$\pm$5.5 \\
\hline
GN1090549 & Extended Emission & 23.1$\pm$6.9 & 31.7$\pm$5.7 & - & - & - & 37.9$\pm$2.6 & 1.5$\pm$5.8 \\
 & Point Source & 8.1$\pm$6.0 & 12.4$\pm$5.1 & - & - & - & 41.7$\pm$12.1 & 146.0$\pm$18.5 \\
\hline
GN1020514 & Extended Emission & 87.1$\pm$5.0 & 49.3$\pm$4.0 & 40.3$\pm$6.4 & 68.8$\pm$4.0 & 72.1$\pm$2.3 & 18.3$\pm$1.8 & 55.1$\pm$2.7 \\
 & Point Source & 26.7$\pm$17.3 & 36.3$\pm$12.1 & 34.3$\pm$21.9 & 35.4$\pm$18.7 & 161.6$\pm$32.6 & 334.7$\pm$15.3 & 399.0$\pm$52.1 \\
\hline
GN1087388 & Extended Emission & 28.1$\pm$15.5 & 25.8$\pm$13.2 & - & - & - & 79.4$\pm$14.7 & 118.6$\pm$25.8 \\
 & Point Source & 16.6$\pm$5.5 & 45.6$\pm$5.2 & - & - & - & 841.2$\pm$11.7 & 2156.7$\pm$22.6 \\
\hline
\enddata
\end{deluxetable*}

\subsection{AGN-Host Spectral Energy Distributions}

Using our best-fit model obtained from GALFIT, we determine the AGN flux contribution by running MCMC. To estimate the host galaxy flux, we apply aperture photometry to the AGN-subtracted image.

We then plot the spectral energy distribution (SED) of all sources. For our sources, we find that for the shorter wavelengths the flux can be dominated either by the extended emission or by the point source. However, we find that for F356W and F444W, over $\sim$50\% and over $\sim$70\% of the light comes from the point source in F356W and F444W. 

Figure~\ref{fig:GN1014406_2D_decomp} (the bottom panel) shows the multi-band SED of GN1014406 as an example. The flux associated with the extended emission remains flat in $f_{\nu}$ across the NIRCam bands, although individual bands may be modestly enhanced by emission lines. The point-source component shows a nearly flat SED in the rest-frame UV and a red slope at LW in the rest-frame optical. In the LW bands, the total flux is dominated by the point-source contribution. 

We conducted SED fitting with \textsc{PROSPECTOR} \citep{Johnson_2021} combined with the Parrot artificial neural network (ANN) emulator for acceleration \citep{Mathews_2023}, following \cite{Zhu}. The source redshift $z_{\rm spec}$ is fixed to the center of the \ha\ narrow component as measured by \citet{Zhang_2025}.

\begin{figure*}[!htbp]
    \centering
    \includegraphics[width=0.9\linewidth]{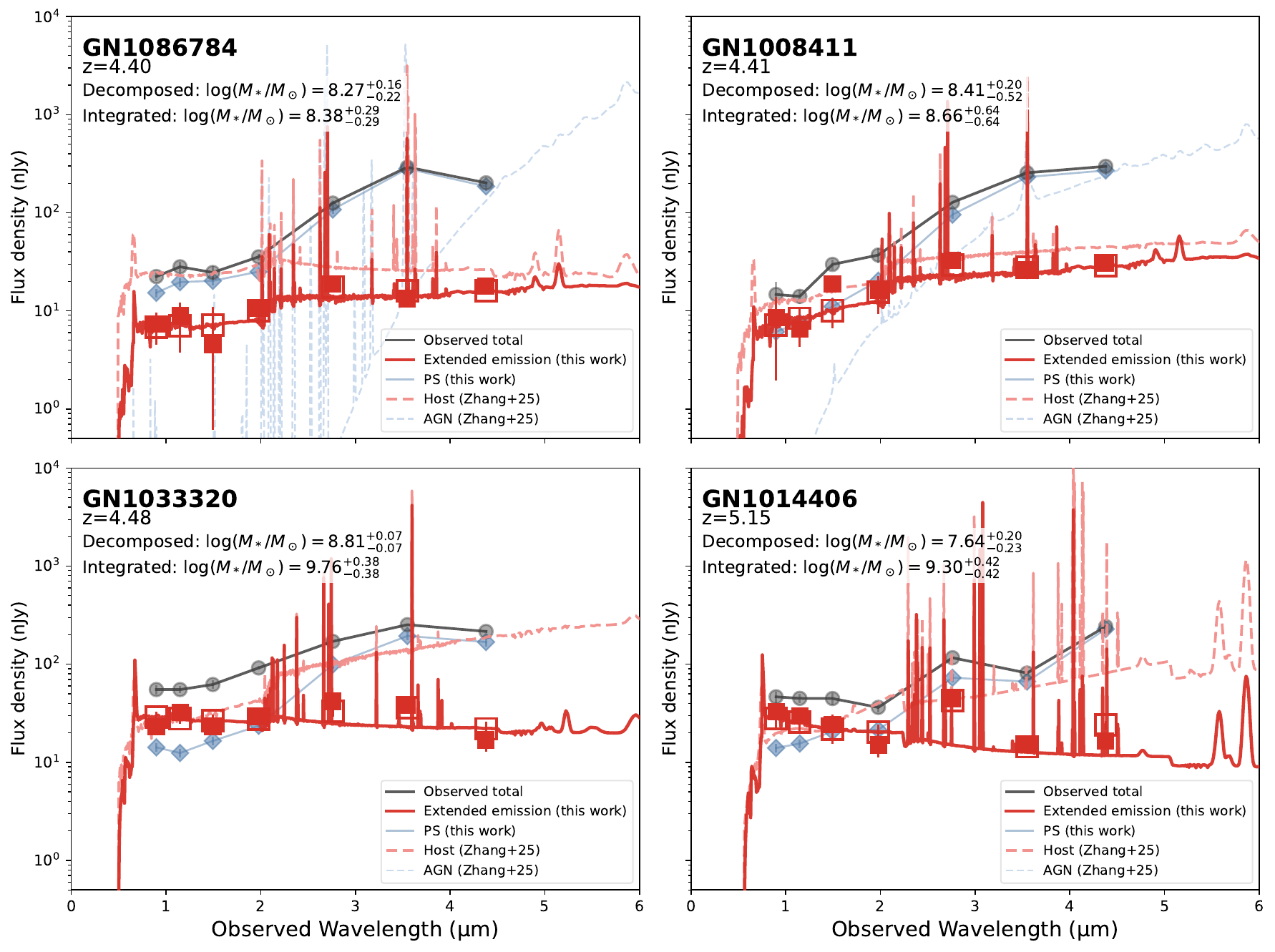}
    \caption{
    SED fits for four selected galaxies. Grey circles show total photometry, and blue diamonds show the decomposed point-source photometry. Solid red lines show host-galaxy SED fits based on image decomposition, while dashed red lines show fits to spatially integrated photometry. Blue dashed lines indicate AGN components inferred from integrated SED fitting \citep{Zhang_2025}.
    {\bf Top two panels:} The SED-fitting analysis infers the presence of an AGN in the spatially-integrated SEDs. The derived host stellar masses are broadly consistent with or without the image decomposition in this case;  {\bf Bottom two panels:} no AGN component was detected without image decomposition. The derived host stellar mass is substantially larger without image decomposition. The full set of fitting results for all sources is available as a figure set and \href{https://github.com/plusante/Decomposition-result}{here}.}
    \label{fig:SED}
\end{figure*}

\citet{Zhang_2025} also performed SED fitting for these sources based on NIRCam$+$HST photometry alone but without image decomposition. Their SED fitting was also performed with \textsc{PROSPECTOR}, but with updated empirical models of the AGN torus emission and hot dust emission from the host galaxy (see \citealp{Lyu_2024} for details). In Figure~\ref{fig:SED}, we compare our SED fitting results based on image decomposition with their results. In our SED fits, which use the decomposed images, we find that the LW fluxes are dominated by the point source in all cases, which we interpret as being due to the AGN. In contrast, \citet{Zhang_2025} find that only half of the sources show AGN evidence from SED fitting. 

\footnote{The figures are also available
\href{https://github.com/plusante/Decomposition-result}{here}.}

We find that the SEDs of our extended emission resemble the underlying stellar host components derived by \citet{Zhang_2025} in the cases where the presence of an AGN is inferred through the SED modeling. In contrast, when the spatially-integrated SEDs show no sign of AGN emission, the fits presented by \citet{Zhang_2025} attribute most of the flux to the host stellar component, exceeding our inferred host emission by a substantial margin.

\section{Discussion}
\label{sec:discussion}

\subsection{The nature of the extended emission} \label{sec:nature}

Our sample is selected as broad \ha \ emitters with ${\rm FWHM_{H\alpha,~broad}}$\,$>$\,1000~\text{km/s} \citep{Zhang_2025}. Such a broad component strongly indicates the existence of an AGN. Based on this assumption, we performed the image decomposition using a PS model to represent the AGN and a S\'ersic model to fit the extended emission.

Here, we examine several interpretations to explain the origin of the residual extended emission. 

{\bf Host-galaxy emission} --- The most likely possibility is that this is the host galaxy \citep{Rinaldi_2025}. This is physically reasonable as when we fit targets with one PSF and one S\'ersic model, the offset between the AGN and S\'ersic centroids is less than 0\farcs06 ( $\lesssim$\,0.36 kpc). Studies of LRD analogs at lower redshift suggest that extended features clearly visible in $z$\,$\sim$\,2 LRDs  would be hard to detect if observed at $z$\,$\sim$\,7 due to cosmological surface brightness dimming \citep{Rinaldi_beyond}.
The fraction of targets in which we detect extended emission ($53 \%$) is consistent with the fraction of LRDs reported to host extended components in previous studies ($\sim 50 \%$, \citealp{Rinaldi_2025}).

Meanwhile, when the extended emission is fitted with galaxy templates, the inferred stellar masses and measured effective radii broadly follow empirical size--mass relations, as discussed in Section~\ref{sec:sizemass}. This further supports our interpretation that the extended emission originates from the host galaxy. 

{\bf AGN-Produced Nebular Emission} --- Another possibility is that the residual signal arises from nebular emission \citep{Chen_2024}. In some LRDs, extended emission has been found to be spatially offset from the AGN component, and in several cases the extended emission is better fit by a pure nebular-emission model \citep{Chen_2025}.

For some sources, however, pure nebular-emission models fail to reproduce the observed fluxes \citep{Chen_2025}. In our sample, we find no obvious offset between the centroid of the point source (i.e., the AGN) and that of the extended emission. Therefore, nebular emission is unlikely to dominate the extended component in our sources..

{\bf Scattered AGN Emission} --- \cite{Labbe_2025} also proposed that the blue rest-frame UV (ultraviolet) light could arise from scattered AGN emission. However, the non-detection of the \HeII\  and \MgII\ emission lines, together with those of the high equivalent-width (EW) \CIII\ and \CIV\, makes this interpretation less likely \citep{Akins_2025}.

For the reasons discussed above, we interpret the extended emission as originating from the host galaxy and perform SED fitting to infer the host-galaxy properties, particularly the stellar masses.

An important caveat of imaging fitting is that, in the absence of spectroscopic redshifts, foreground galaxies projected along the line of sight may be mistakenly identified as physically associated close companions.
Two sources in our sample (GN1033320, GN1029154) where the residual light  that initially appeared to arise from companion galaxies. After examining their spectra, we found that: (1) both objects are actually foreground interlopers; (2) their morphologies resemble off-centered nebular emission; and (3) without spectroscopic confirmation, they can be easily misinterpreted as companions or nebular emission. This highlights that imaging alone could be misleading because nebular emission and foreground galaxies can mimic host features.

\begingroup
\setlength{\tabcolsep}{15pt}
\begin{deluxetable*}{lcccc}
\tablewidth{0pt}
\label{table:mstar}
\caption{Black-hole and stellar masses of the 15 AGN galaxies with host detection} 
\tablehead{
  \colhead{JADES ID} & \colhead{$\log(M_{\rm BH}/M_{\odot})$} & \colhead{$\log{(M_{*,\,\rm decom.}/M_{\odot})}$} & \colhead{$\log{(M_{*,\,\rm SED}/M_{\odot})}$} & \colhead{AGN Evidence}
}
\decimalcolnumbers
\startdata
GN1087315 & 7.01 $\pm$ 0.11 & $7.89^{+0.10}_{-0.11}$ & $8.76 \pm 0.28$ & N \\
GN1082263 & 6.65 $\pm$ 0.17 & $8.58^{+0.08}_{-0.15}$ & $8.77 \pm 0.40$ & Y  \\
GN1089568 & 7.04 $\pm$ 0.09 & $8.91^{+0.07}_{-0.24}$ & $8.76 \pm 0.25$ & Y \\
GN1029154 & 7.38 $\pm$ 0.10 & $8.65^{+0.17}_{-0.42}$ & $8.12 \pm 0.32$ & Y \\
GN1088832 & 7.85 $\pm$ 0.04 & $9.22^{+0.11}_{-0.25}$ & $9.64 \pm 0.21$ & Y \\
GN1086784 & 7.71 $\pm$ 0.15 & $8.27^{+0.16}_{-0.22}$ & $8.38 \pm 0.29$ & Y\\
GN1086855 & 7.28 $\pm$ 0.14 & $8.32^{+0.01}_{-0.01}$ & $8.85 \pm 0.43$ & Y \\
GN1008671 & 7.59 $\pm$ 0.04 & $8.78^{+0.16}_{-0.05}$ & $10.14 \pm 0.46$ & N\\
GN1008411 & 7.72 $\pm$ 0.21 & $8.41^{+0.20}_{-0.52}$ & $8.66 \pm 0.64$ & Y\\
GN1033320 & 7.22 $\pm$ 0.19 & $8.81^{+0.07}_{-0.07}$ & $9.76 \pm 0.38$ & N \\
GN1090253 & 7.13 $\pm$ 0.07 & $9.17^{+0.01}_{-0.01}$ & $9.31 \pm 0.50$ & Y\\
GN1014406 & 7.81 $\pm$ 0.18 & $7.64^{+0.20}_{-0.23}$ & $9.30 \pm 0.42$ & N\\
GN1090549 & 7.19 $\pm$ 0.16 & $8.73^{+0.08}_{-0.10}$ & $9.09 \pm 0.19$ & Y\\
GN1020514 & 7.22 $\pm$ 0.08 & $8.32^{+0.08}_{-0.08}$ & $10.14 \pm 0.42$ & N\\
GN1087388 & 8.29 $\pm$ 0.02 & $9.90^{+0.23}_{-0.45}$ & $10.87 \pm 0.26$ &N \\
\enddata
\tablecomments{Columns: (1) JADES IDs of the AGN. (2) Black Hole masses reported by \cite{Zhang_2025}. (3) Stellar mass estimates derived from SED fitting based on imaging decomposition results (this work). (4) Stellar mass estimates reported by \cite{Zhang_2025}, from photometry-based SED fitting. (5) Whether AGN evidence is prominent in photometry-based SED fitting \citep{Zhang_2025}.}
\end{deluxetable*}
\endgroup

\subsection{Stellar Mass: With/Without Image Decomposition}
\label{sec:stellar_mass}

\begin{figure}[!ht]
\centering
\includegraphics[width=\linewidth]{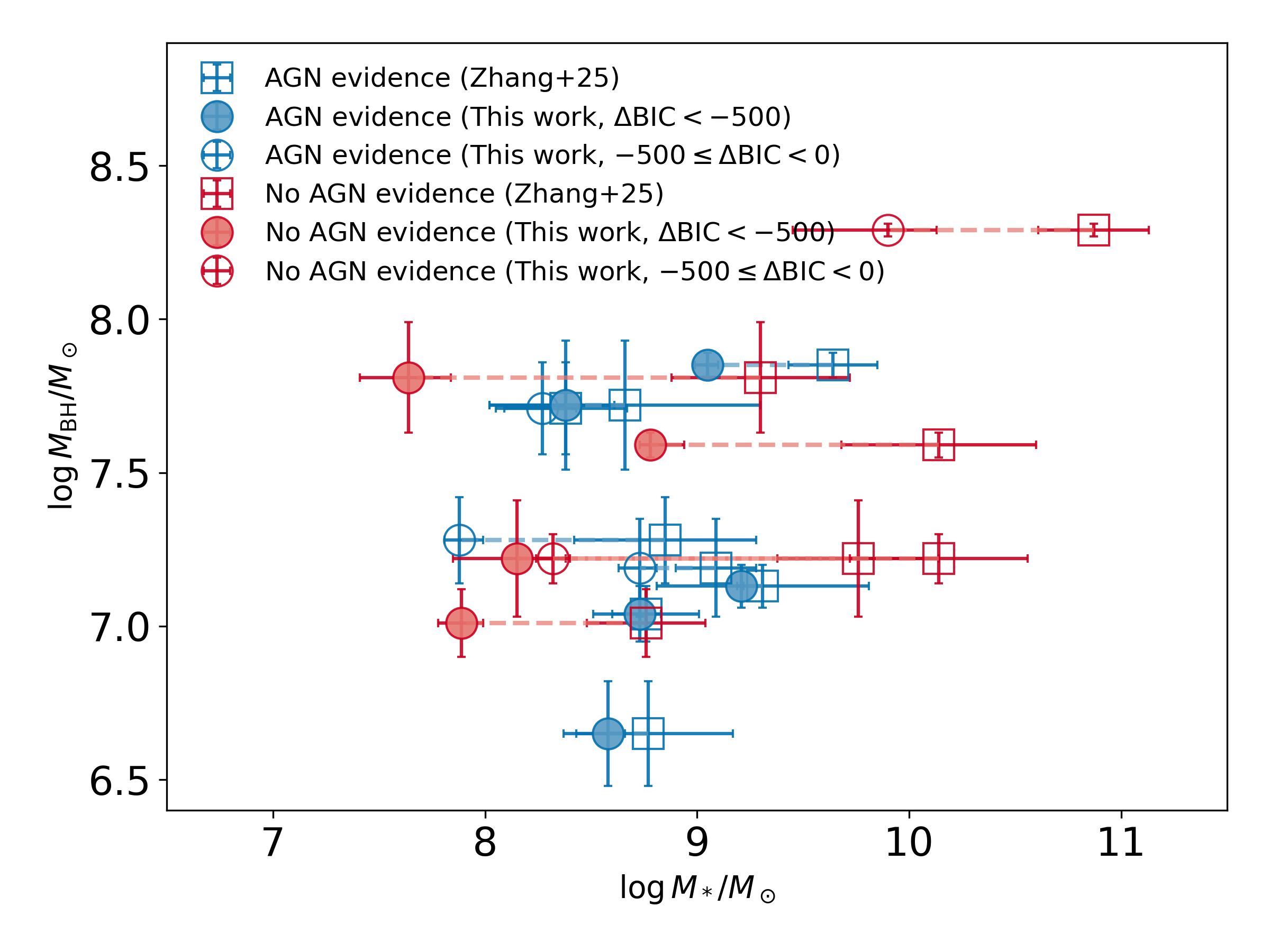}
\caption{Comparison of stellar mass estimates. Stellar masses derived from our analysis are shown as circles, while those from \cite{Zhang_2025} are shown as squares. Measurements for the same sources are connected by dashed lines. Galaxies with AGN evidence in their spatially-integrated SEDs are shown in blue, while sources without AGN evidence in their spatially integrated SEDs are shown in red. Spatially integrated SED fitting can overestimate stellar masses by 1 -- 2 dex when AGN component is not detected.}
\label{fig:mass_compare_with_junyu}
\end{figure}

We list the stellar masses for all targets with detected hosts in Table \ref{table:mstar}, along with the black hole and stellar masses reported by \citet{Zhang_2025}. In Figure~\ref{fig:mass_compare_with_junyu}, we compare our stellar‐mass measurements with those reported in \citet{Zhang_2025}.

For targets in which an AGN component is identified in SED fitting of \cite{Zhang_2025}, the SED of the host galaxy derived from our image decomposition closely resembles the stellar component SED in their results. Consequently, we find consistent stellar mass estimates, with differences typically $\lesssim 0.5$~dex. 

In contrast, for sources without evidence for an AGN component in their photometry-based SED fitting, the inferred host-galaxy SED from our analysis differs significantly from theirs, leading to substantially different stellar mass estimates. In these cases, our stellar masses are $\sim 0.9$--$1.8$~dex lower than those reported in \cite{Zhang_2025}.

These discrepancies arise from differences in how flux is attributed between the stellar and AGN components. SED fitting based solely on spatially-integrated photometry may be unable to robustly separate multiple components. In the cases where the SED fit correctly identifies a non-negligible AGN component, the inferred stellar masses are broadly consistent with those obtained when the host and AGN are separated via image decomposition. However, when the AGN contribution is not recognized (or is weak enough to be absorbed by other SED-fitting degrees of freedom), a substantial fraction of the flux can be incorrectly assigned to the stellar component, leading to an overestimation of the stellar mass by $\sim$1--2~dex.

Similar conclusions were reached by \cite{Pablo_2024}, who showed that the stellar masses inferred from spatially integrated SED fitting depend strongly on the assumed AGN contribution. Specifically, stellar masses can be more than 1 dex higher when the AGN contribution is left unconstrained than when the AGN is forced to dominate the rest-frame optical and infrared emission. However, in the absence of imaging-based decomposition, there is no strong justification for constraining the relative contributions of the stellar and AGN components rather than allowing them to vary freely, especially when mid-infrared data are missing. This comparison highlights that neglecting imaging-based decomposition can significantly affect SED-derived stellar masses, particularly in the systems with uncertain AGN contributions.

It has also been reported that SED fitting based on integrated NIRCam photometry may overestimate stellar masses when longer-wavelength data are unavailable, likely due to weak constraints on the rest-frame optical continuum \citep{Christina2024, Giovanni_2026}.
Including MIRI and ALMA data can lead to a $\sim$ 0.6 dex decrease in stellar mass estimation compared to fits based only on HST + NIRCam data even assuming pure stellar origin \citep{Christina2024, Wang_2025}.  

A caveat of our analysis is that, although we have attempted to separate the AGN and host flux through PS+S\'ersic decomposition, we cannot rule out the possibility that some fraction of the host mass is highly concentrated in the central region and therefore cannot be separated from the point-like AGN emission, even at the JWST/NIRCam spatial resolution. As tested in Appendix~\ref{flux-recovery}, compact S\'ersic radial-profile can be recovered as long as they are not extremely compact relative to the PSF. If such an unresolved central component exists, then these host galaxies could have an extremely concentrated mass distribution, while still showing some extended emission in the SW bands. In either case, whether the high $M_{\text{BH}}/M_\ast$ ratios revealed by our image decomposition (and by previous work) are real, or the host galaxies are so compact that part of their mass cannot be separated from the AGN, the results support an inside-out growth picture for these AGN-host systems.  

\subsection{Size-Mass Relation}
\label{sec:sizemass}
\begin{figure}[!ht]
\centering
\includegraphics[width=\linewidth]{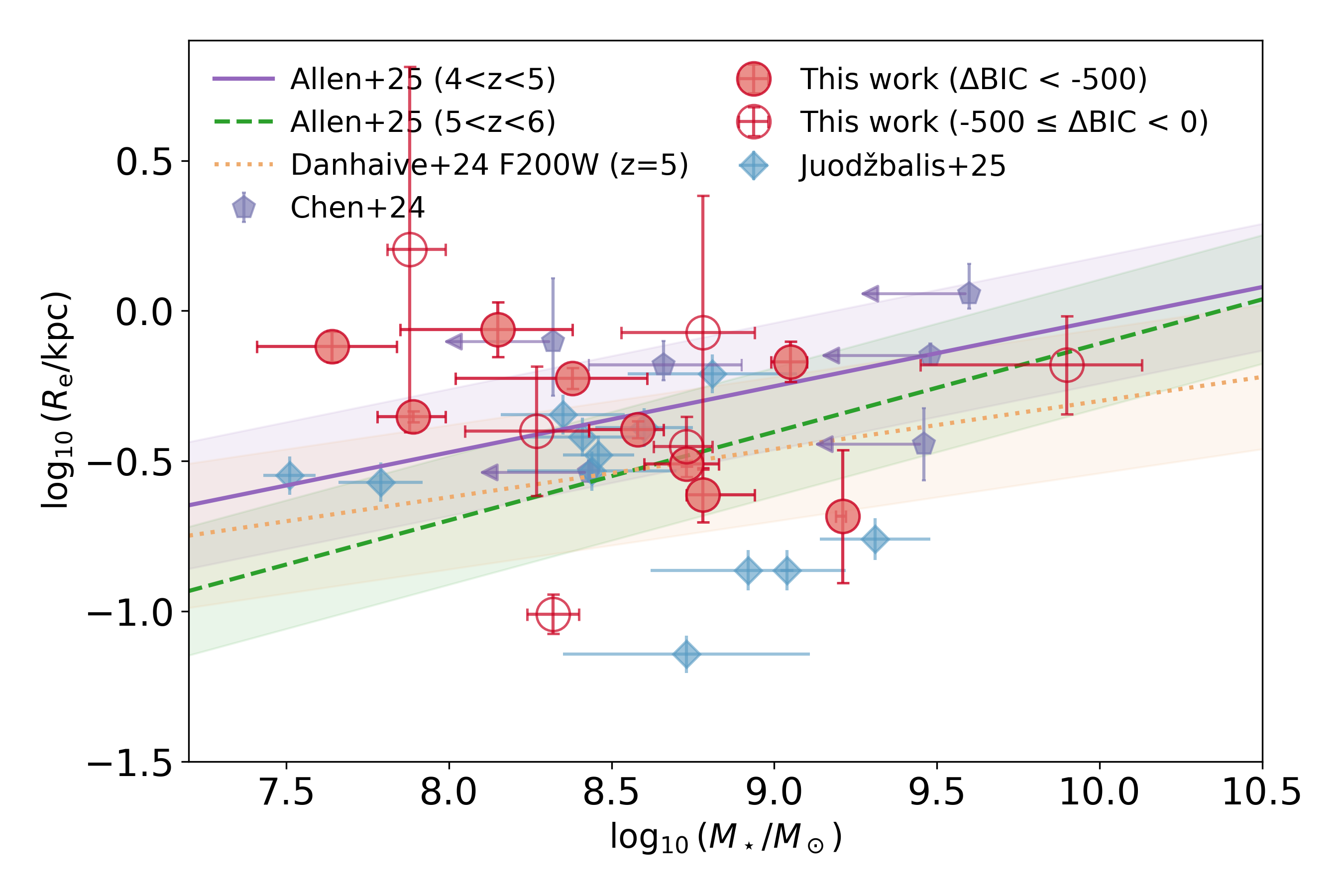}
\caption{Stellar mass--size relation for our sample. Effective radii obtained from imaging-based AGN--host decomposition are shown as red circles. Filled red circles indicate sources with $\Delta{\rm BIC} < -500$, while open red circles indicate sources with $-500 \leq \Delta{\rm BIC} < 0$. The purple solid line and green dashed line with shaded regions represent the best-fit size--mass relations and intrinsic scatter from \citet{Allen_2025} in the corresponding redshift bins. The orange dotted line shows the size--mass relation from \citet{Lola_2025}. Measurements from \citet{Chen_2024} and \citet{Ignas_2025} are included as purple pentagons and blue diamonds for comparison. The sizes and stellar masses of our sources follow the empirical relations, supporting the interpretation that the extended component recovered in our decomposition corresponds to the stellar component of the host galaxies.}
\label{fig:sizemass}
\end{figure}

In Figure~\ref{fig:sizemass}, we present the stellar mass--size relation for our sample, overplotted with the size--mass relations from \cite{Allen_2025} and \cite{Lola_2025}. For our analysis, we adopt the S\'ersic model derived from a representative SW band, which can range from F090W to F200W, but is most commonly selected from F115W or F150W. To enable a consistent comparison, we therefore use the F150W size--mass relation from \cite{Allen_2025} and F200W size--mass relation from \cite{Lola_2025}. Given the small sample size and the limited spatial resolution imposed by the PSF size, our data exhibit slight deviations from the empirical relation; however, most data points remain within the ±1σ intrinsic scatter.
We also include measurements from \cite{Chen_2024}, whose size--mass relation is broadly consistent with our results. 
Figure~\ref{fig:sizemass} further supports our assumption that the extended emission is likely from the host galaxy.  

\subsection{\texorpdfstring{$\text{M}_{\text{BH}}- \text{M}_*$}{MBH-M*} Relation}

\begin{figure*}[!ht]
\centering
\includegraphics[width=\linewidth]{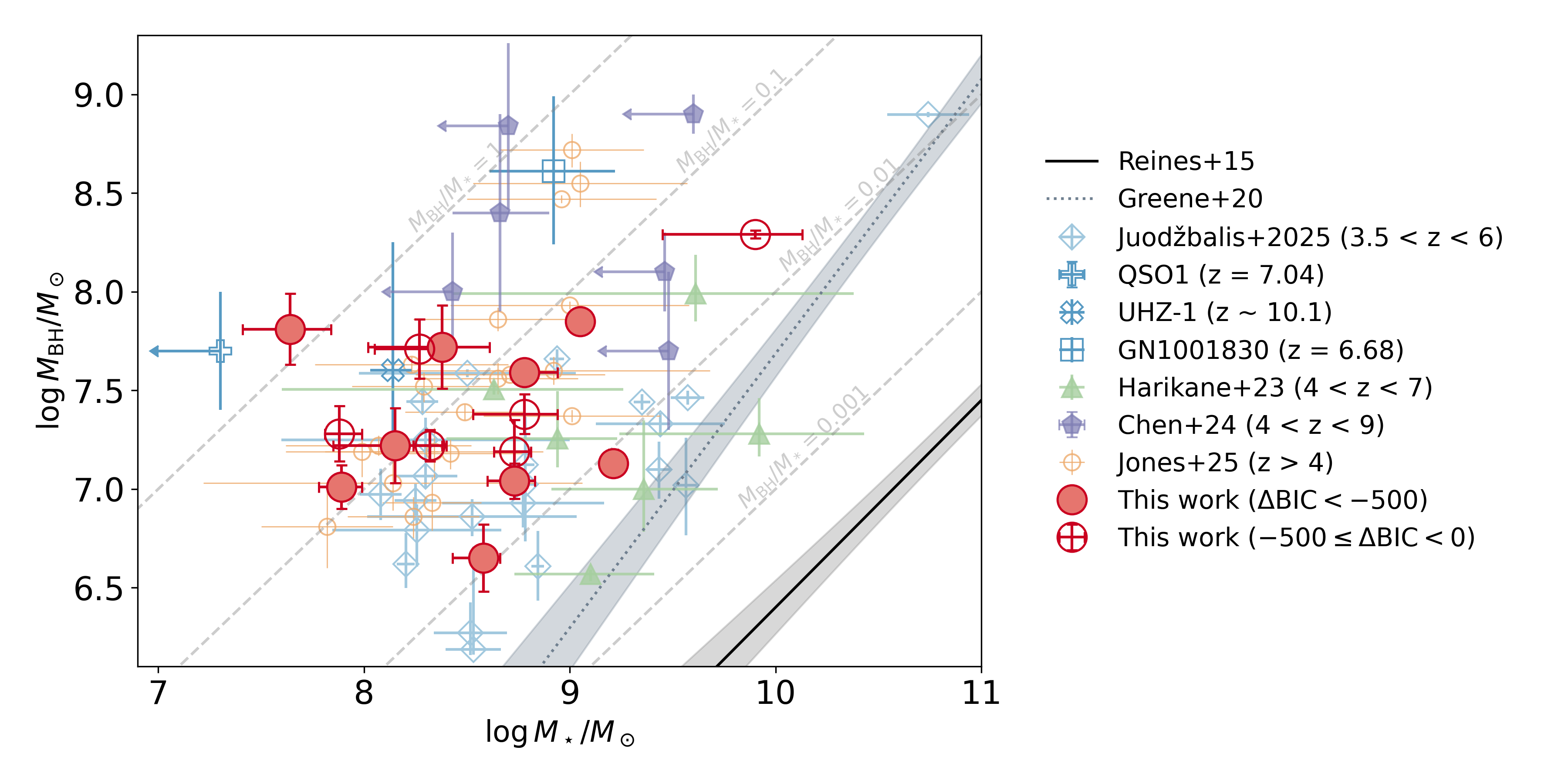}
\caption{BH mass versus host galaxy stellar mass. Targets for which we detect emission from the host galaxy are shown as red solid circles. Green triangles are measurements from \cite{harikane_23}, where they decomposed the AGN/Host galaxy for faint type-1 AGN using JWST/NIRCam and HST data. Purple pentagons show measurements of broad-line AGN from  \cite{Chen_2024} based on image decomposition. Light-blue diamonds are measurements from \cite{Ignas_2025} where stellar masses were estimated with SED fitting based on total photometry. UHZ-1, QSO1, and GN1001830 are shown as a blue cross, a plus symbol, and a square, respectively. Orange circles are stellar mass estimates of LRDs from \cite{Jones_2025}.
The solid black and dotted gray lines are the local $\text{M}_{\text{BH}}- \text{M}_*$ relation inferred by \cite{Reines_2015} and \cite{Greene_2020}, with uncertainties represented by the shaded regions.  Gray dashed lines correspond to $\text{M}_{\text{BH}}/\text{M}_*$ = 1, 0.1, 0.01 and 0.001. Our sources show an elevated $\text{M}_{\text{BH}}- \text{M}_*$ relation compared to the local relations, consistent with results from other faint AGN surveys.}
\label{fig:mbh_mstar}
\end{figure*}

Figure~\ref{fig:mbh_mstar} shows the inferred stellar and black hole masses for our sources, together with results from other studies. We compare our measurements to AGN of similar black hole masses identified in recent JWST surveys \citep{harikane_23, maiolino_24, Bogdan_24}, and to the local M$_{\text{BH}}$/M$_*$ relation reported by \citet{Reines_2015} and \citet{Greene_2020}.

For objects with detected host-galaxy emission, their black holes appear extremely overmassive relative to the host galaxies when compared to the local relation of \citet{Reines_2015}. When compared to the local relation from \citet{Greene_2020}, which is suggested to be more robust and less sensitive to galaxy type \citep{Sun_2025}, the inferred M$_{\text{BH}}$/M$_*$ ratios lie closer to the local relation but are still well above it.

The stellar masses derived for our sources are typically a factor of $\sim100$--1000 lower than expected from the local $\mathrm{M}{_\mathrm{BH}}$--$\mathrm{M}_*$ relation. This extreme offset suggests that black holes in these early AGN may grow more rapidly than their host galaxies, or that star formation in the host galaxies is significantly suppressed, potentially by AGN feedback. Many recently discovered faint AGN at high redshift have also been reported to lie above the local relation, hosting black holes that are overmassive relative to their stellar mass; for our sample, this discrepancy appears even more pronounced.

Our M${_\text{BH}}$/M$_*$ ratios span 0.01--1.48. This range is broadly consistent with the results based on image decomposition for LRDs (0.02--1.38; \citealp{Chen_2024}). For comparison, faint AGN galaxies analyzed using NIRSpec prism spectra and multiband photometry yield lower ratios of 0.002--0.22 \citep{Ignas_2025}. The M${_\text{BH}}$/M$_*$ ratios of LRDs from \cite{Jones_2025} is inferred from SED-fitting alone as well. Although they performed image decomposition using NIRCam imaging data, this decomposition was used only to assess whether extended emission is present. The total flux was still taken as the input for SED fitting, with the flux attribution between the AGN and the host galaxy determined entirely by the SED fitting.

This difference is expected. As we have reporteded in Section~\ref{sec:stellar_mass}, when we compare our stellar mass estimates to those obtained from the SED fitting with spatially-integrated photometry, we find that the latter can be 1--2 dex higher, especially when AGN emission is not strongly evident in the SED fitting based on total photometry. In such cases, the stellar continuum may be overestimated, leading to larger inferred stellar masses and lower M${_\text{BH}}$/M$_*$ ratios. A similar effect is also reported in \cite{Ignas_2025}, where stellar mass estimates derived from image decomposition are lower than those obtained from spectra-based SED fitting by up to $\sim$2.5 dex, and lower than those from SED fitting based on total photometry by up to $\sim$1.2 dex.

We also include AGN galaxies from \cite{harikane_23}, where stellar masses are also derived using image decomposition. Their sample selection differs from ours: they did not restrict the AGN hosts to be spatially compact, leading to systematically higher stellar masses in their sample.

Some of our sources have extreme M$_{\text{BH}}$/M$_*$ ratios above 0.1. There are several sources reported in literature having extreme M$_{\text{BH}}$/M$_*$ ratios as well, like UHZ1 \citep{Bogdan_24, Goulding_23}, QSO1 \citep{Furtak_QSO1, Furtak_2024, DEugenio_QSO1, Ignas_2025_bhmass}, GN1001830 \citep{Ignas_dormant}, suggesting that it may be a common feature for supermassive black holes (SMBHs) to be significantly overmassive with respect to their host galaxies in the early universe. The high BH mass and high ratio of M$_{\text{BH}}$/M$_*$ of these sources are used to support the heavy black hole seeding scenario, such as Direc Collapse Black Holes \citep{Natarajan_2024, Ignas_2025_bhmass}. The similarly elevated $M_{\rm BH}/M_*$ ratios in our sources may indicate suppressed or delayed stellar mass assembly in their host galaxies—potentially due to AGN feedback—or alternatively reflect rapid early SMBH growth or a heavy black hole seeding scenario.

\subsection{Is the Broad-\ha-Derived $M_{\rm BH}$ Really Reliable?}

The black hole masses we used here are measured by \cite{Zhang_2025}, who used single-epoch virial estimators proposed by \citet{Reines_2013},
\begin{equation}
\begin{split}
    {\rm log_{10}(M_{\rm BH}/M_{\odot})} = 6.57 + {\rm log}_{10}(\epsilon) + \\ 0.47 \, {\rm log}_{10}(L_{\rm H\alpha, broad}/10^{42}\, {\rm erg \,s}^{-1}) + \\ 2.06 \,  {\rm log}_{10}({\rm FWHM_{broad}}/10^3\, {\rm km \, s^{-1}}).
\end{split}
\end{equation}
This approach assumes the  broadening of \ha \ is entirely due to Doppler motions. However, the applicability of single-epoch virial mass estimators at high redshift remains uncertain, as there are currently no direct calibration measurements beyond $z \gtrsim 2$. \citet{Rusakov_2025} proposed that little red dots (LRDs) may host young supermassive black holes embedded in dense ionized gas, where emission lines are significantly broadened by electron scattering rather than purely by virial motions. Under this scenario, the inferred black-hole masses are 2 dex lower than those derived with only Doppler broadening. If correct, this would substantially alleviate the apparent tension between the high $M_{\rm BH}/M_\ast$ ratios observed at high redshift and local scaling relations. Yet, a direct, dynamical black hole mass measurement for an LRD results in a value consistent with those obtained from single epoch virial relation \citep{Ignas_2025_bhmass}, suggesting that black holes in faint high-redshift galaxies may indeed be intrinsically overmassive.

\section{Conclusion}\label{sec:conclusion}

In this study, we performed spatially resolved AGN--host galaxy decomposition for a sample of 17 broad-line AGN galaxies at $z \sim 5$ using deep JWST/NIRCam imaging from the JADES survey. Our sample consists of  broad \ha\ emitters at $z = 3.7$--6.5 identified in the CONGRESS and FRESCO surveys over the JADES GOODS-N field. The image decomposition is carried out using JADES NIRCam observations in up to seven wide bands (F090W--F444W).

We use a two-component model consisting of a S\'ersic profile to represent the host galaxy and a point source to represent the AGN. Structural parameters of the S\'ersic component are determined using \textsc{GALFIT}, while the flux contributions of the AGN and host galaxy in each band are estimated using MCMC fitting.

Our key results are as follows:

\begin{itemize}
    \item Among the 17 broad-line AGN galaxies in our sample, we detect extended host-galaxy emission in 9 galaxies. For 6 of the galaxies with no detection, the fit does not show a significant improvement when using a two-component (PS + S\'ersic) model compared to a single PS model ($-$500\,$<$\,$\Delta{\rm BIC}$\,$<$\,0). For the remaining 2 galaxies with no detection, a single PS model is formally preferred ($\Delta{\rm BIC}$\,$>$\,0).
    The radial profiles of these two galaxies in both short-wavelength (SW) and long-wavelength (LW) bands closely follow the PSF, indicating that their emission is AGN-dominated across all bands.
    
    \item  We find that incorporating image decomposition into SED fitting can lead to stellar mass estimates 1--2 dex lower than those obtained without image decomposition {\em when the AGN component is not detected in the sptially-integrated SEDs}. In the cases where clear AGN signatures are already present in the spatially-integrated SEDs, SED-fitting analyses with or without image decomposition would produce consistent stellar masses, typically differing by $\lesssim 0.6$ dex.
    
    \item The detected host galaxies exhibit relatively low stellar masses compared to their black hole masses, resulting in a significant offset from the local M$_{\text{BH}}$/M$_*$ scaling relation. Elevated M$_{\text{BH}}$/M$_*$  ratios have also been reported in other LRD surveys; however, our galaxies show systematically higher ratios compared to the results based on the SED fitting without image decomposition. In contrast, when compared to LRDs whose stellar masses are derived using image decomposition, our inferred ratios are consistent.

\end{itemize}

We demonstrate that extended emission can be robustly identified in $\sim$50\% of galaxies hosting faint AGN. By performing imaging-based AGN--host decomposition, we recover the host-galaxy's SED and from the latter derive its physical properties.  Our results highlight the importance of spatially separating AGN and host emission when characterizing the stellar content of faint AGN populations. Future work incorporating spatially-resolved spectroscopic observations will allow us to directly link host-galaxy structure to AGN activity, providing deeper insight into the co-evolution of faint AGN and their host galaxies.

\begin{acknowledgments}
ZM, EE, YZ, and CNAW acknowledge support from the NIRCam Science Team contract to the University of Arizona, NAS5-02105. YZ is also supported by JWST Program \#6434. Support for program \#6434 was provided by NASA through a grant from the Space Telescope Science Institute, which is operated by the Association of Universities for Research in Astronomy, Inc., under NASA contract NAS 5-03127. AJB acknowledges funding from the “FirstGalaxies” Advanced Grant from the European Research Council (ERC) under the European Union’s Horizon 2020 research and innovation program (Grant agreement No. 789056). SC acknowledges support by European Union’s HE ERC Starting Grant No. 101040227 - WINGS. ECL acknowledges support of an STFC Webb Fellowship (ST/W001438/1). RM acknowledges support by the Science and Technology Facilities Council (STFC), by the ERC through Advanced Grant 695671 “QUENCH”, and by the UKRI Frontier Research grant RISEandFALL. RM also acknowledges funding from a research professorship from the Royal Society. XJ acknowledges support by the Science and Technology Facilities Council (STFC), by the ERC through Advanced Grant 695671 ``QUENCH'', and by the UKRI Frontier Research grant RISEandFALL. ST acknowledges support by the Royal Society Research Grant G125142. H\"U acknowledges funding by the European Union (ERC APEX, 101164796). Views and opinions expressed are however those of the authors only and do not necessarily reflect those of the European Union or the European Research Council Executive Agency. Neither the European Union nor the granting authority can be held responsible for them.

This work is based on observations made with the NASA/ESA/CSA James Webb Space Telescope. The data were obtained from the Mikulski Archive for Space Telescopes at the Space Telescope Science Institute, which is operated by the Association of Universities for Research in Astronomy, Inc., under NASA contract NAS 5-03127 for JWST.

The authors acknowledge the FRESCO team for developing their observing program with a zero-exclusive-access period.

All the JWST data used in this paper can be found in MAST: \dataset[https://doi.org/10.17909/8tdj-8n28]{https://doi.org/10.17909/8tdj-8n28}.

This material is based upon High Performance Computing (HPC) resources supported by the University of Arizona TRIF, UITS, and Research, Innovation, and Impact (RII) and maintained by the UArizona Research Technologies department. This project made use of lux supercomputer at UC Santa Cruz, funded by NSF MRI grant AST 1828315. 

We respectfully acknowledge the University of Arizona is on the land and territories of Indigenous peoples. Today, Arizona is home to 22 federally recognized tribes, with Tucson being home to the O’odham and the Yaqui. The university strives to build sustainable relationships with sovereign Native Nations and Indigenous communities through education offerings, partnerships, and community service.

\end{acknowledgments}

\vspace{5mm}
\facilities{JWST, MAST}

\software{
{\tt GALFIT} \citep{peng2002, peng2010}
{\tt PROSPECTOR \citep{Johnson_2021}}
}

\clearpage

\appendix

\section{Testing the Robustness of Host Galaxy Flux Recovery}
\label{flux-recovery}

Separating the emission from an AGN and its host galaxy has long been a difficult task, particularly at high redshift where quasar hosts are extremely compact and the AGN overwhelmingly dominates the observed light \citep{Ding_2023}. Although the AGN in our sample are relatively faint---mitigating this issue to some extent---the decomposition remains non-trivial. When \textsc{GALFIT} returns a very compact S\'ersic component, we are especially concerned about degeneracies between this compact S\'ersic model and the point source model.

Recovering the host galaxy flux in the LW bands is even more challenging. In F356W and F444W, the NIRCam images of our targets appear unresolved, and for systems with host detections, the AGN contributes more than 50\% of the total light. The radial flux profiles in these bands closely resemble those of a point source, indicating that the host galaxy is strongly outshined by the AGN. In addition, the LW PSFs are intrinsically broader than those in the SW channels, further increasing the degeneracy between the AGN and host components. Consequently, a fraction of the stellar emission may be misassigned to the AGN during decomposition, or conversely, AGN light may be incorrectly attributed to the host.

To assess the reliability of our host flux measurements , we performed a series of mock experiments. We selected a compact galaxy at $z = 4.83$ in GOODS-N field, GN1034159, from the JADES NIRCam and NIRSpec catalogs. We first fit this galaxy with one point source and one S\'ersic model, the same procedure employed for our study as described in this paper. The fit using F115W gives an estimate of Re = 0.38 kpc and n = 1.46. These values are similar to the best-fit models we derived for the host galaxies of our sample, so we believe that this galaxy can represent the host galaxy in our sample well. Using this galaxy as a template for the host galaxies, we constructed mock galaxy images to test AGN-host decomposition performance. 

\subsection{The effects of AGN-host flux ratios}

We first test how the AGN-host flux ratio may affect the recovery of the host flux. To create the mock data, we added scaled AGN (i.e., PS) model images on top of each galaxy image. Each AGN component was scaled such that the AGN-to-host flux ratio takes values in the range from 0 to 100000, producing a series of synthetic AGN-contaminated galaxies. We included the Poisson noise from the AGN by adding its contribution to the error map and injecting a PS model with the corresponding random Poisson noise into the science image to account for the additional noise expected when a real AGN is present.

We fit the real galaxy images with \text{GALFIT}, using a composite S\'ersic$+$PS model. The best-fit S\'ersic model from the band with the most reliable fit was then taken as the representative host model. This representative model was subsequently convolved with the PSF in each band to generate host models. Together with the PS model, these S\'ersic models were used as inputs to our MCMC fitting procedure, which we applied to each mock image to determine the recovered host and AGN fluxes.

\begin{figure}[!ht]
\centering
    \includegraphics[width=\linewidth]{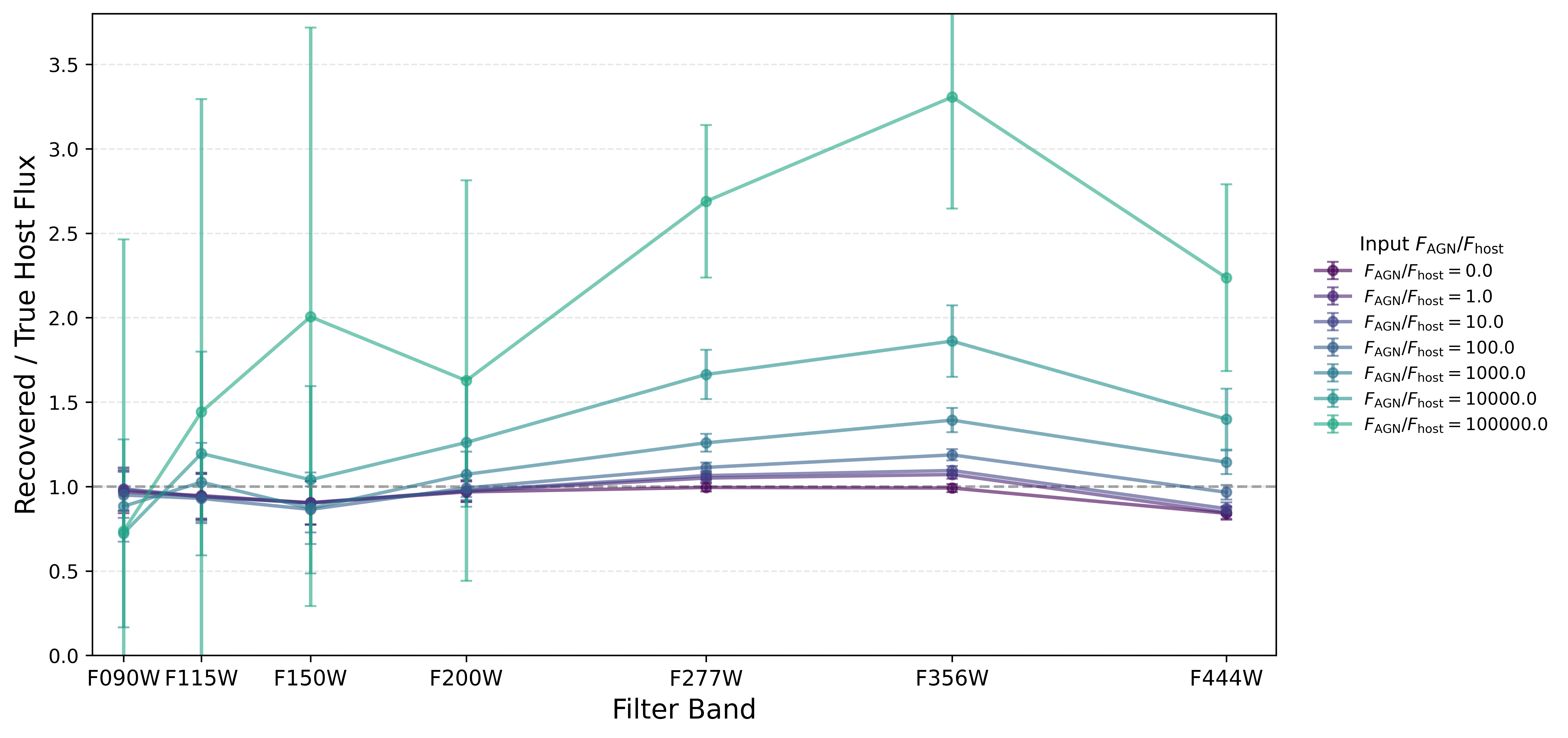}
\hfill
\caption{Ratio of recovered host fluxes to real host fluxes. Each color denotes a different injected AGN-to-host flux ratio (legend). Recovered host fluxes are consistent with the true values within $\sim$ 10 \% across most bands when AGN/host ratio is $<$ 10. At higher AGN/host ratios the scatter increases, yet the decomposition remains stable even for AGN-dominated cases up to a ratio of 20. Error bars indicate $1\sigma$ uncertainties. The dashed line represents the ideal case in which the host flux is correctly recovered (i.e., recovered/true\,$=$\,1).}
\label{fig:host_recover_ratio}
\end{figure}

We compared the recovered host fluxes with the true host fluxes measured through aperture photometry on the original galaxy images. The ratio of recovered to true fluxes in each band is shown in Figure~\ref{fig:host_recover_ratio}. Even when no AGN component is added, fitting a PS model plus a S\'ersic model introduces a systematic bias---particularly in the F444W band---because the model is forced to distribute the central light between the S\'ersic and PS components, leading to a fraction of the host flux being misassigned to the PS model. In real galaxies, any intrinsically compact central light concentration could therefore be incorrectly interpreted as an AGN.

The recovery of host--galaxy fluxes depends strongly on the AGN/host ratio.
As the AGN contribution increases, the decomposition becomes more degenerate, which causes the recovered host flux to deviate more from the true value.

From F090W to F200W, the difference between the injected and recovered host flux stays below 15\% for AGN/host ratios up to $\sim$ 1000. In F277W, the recovery remains accurate (within 15\%) only when the AGN/host ratio is below $\sim$200. At longer wavelengths, the breakdown occurs at even lower AGN/host ratios. In F356W, reliable recovery (difference $<$ 15\%) is maintained only for AGN/host ratios below 40.

F444W is more challenging.
Due to the poor spatial resolution and stronger PSF--Sérsic degeneracy, even at AGN/host ratio = 0 the recovered host flux reaches only 84.4\% of the true value. This intrinsic bias reflects the limited ability to distinguish a compact host from the PSF in the longest--wavelength band. This demonstrates that the ability to recover the host flux is strongly dependent on the resolution of the filter.

\subsection{S\'ersic Profile Recovery}
\label{sec:sersic_profile_recovery}
We broadly recover the host-galaxy flux in all bands ($\lesssim$\,40\%) even for AGN-to-host flux ratios up to 1000. The S\'ersic structural parameters are derived from the original (non-AGN-injected) F115W image. Applying this fixed structural model to other bands enables robust host-flux recovery, even under extreme AGN dominance.

\begin{figure}[!ht]
\centering
    \includegraphics[width=\linewidth]{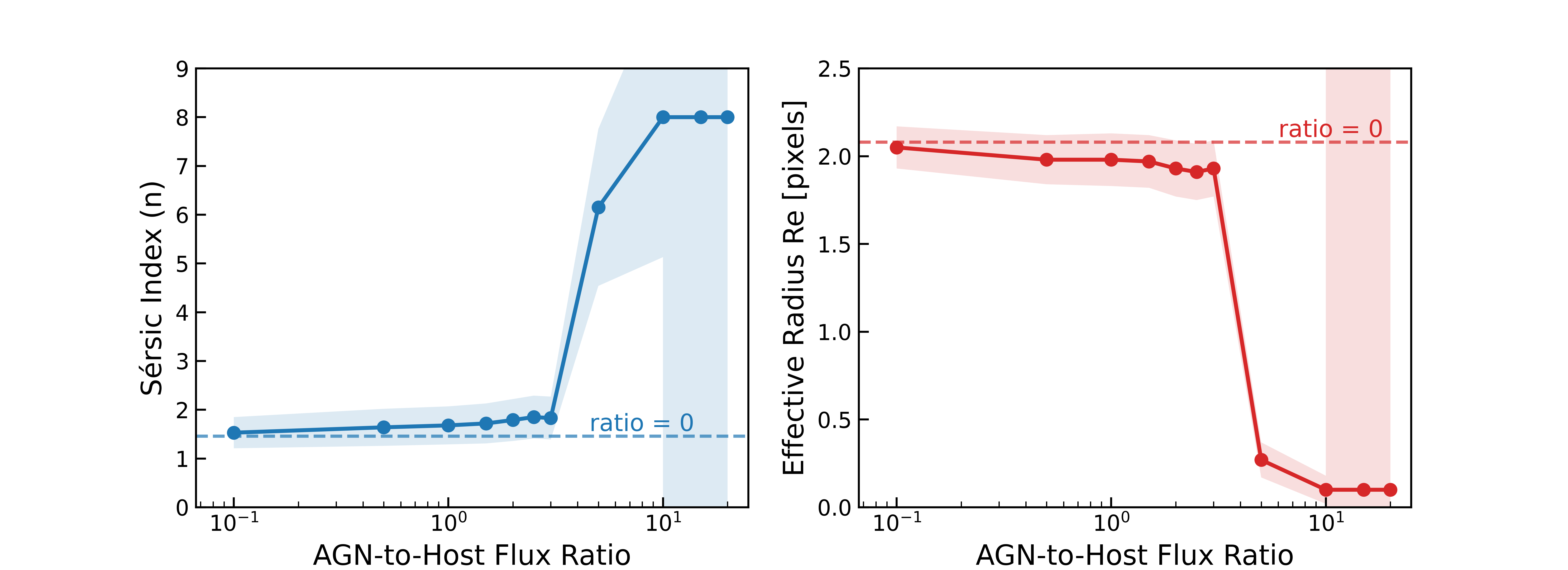}
\hfill
\caption{S\'ersic parameter recovery as a function of AGN-to-host flux ratio in the F115W band.
The left panel shows the recovered S\'ersic index 
n, and the right panel shows the effective radius $R_e$ in pixels, both derived from the \textsc{galfit} decomposition of mock AGN--host images. The S\'ersic parameters obtained from the original galaxy image (with no injected AGN component) are shown as dashed lines. The shaded regions indicate the $1\sigma$ uncertainties reported by \textsc{galfit}.}
\label{fig:params_recovery}
\end{figure}

Now we tested the recovery of the S\'ersic profile parameters under different injected AGN-to-host flux ratios to assess when reliable host-galaxy structural measurements can be obtained. For each ratio, we fit the mock images using \textsc{galfit} and take the S\'ersic parameters recovered from the F115W band as the model of the host galaxy. Fig~\ref{fig:params_recovery} shows how the recovered S\'ersic parameters vary with flux ratio. The recovered parameters remain stable and consistent with the input model up to AGN-to-host ratios of $<$3 . At ratios of 5, the recovered parameters begin to deviate noticeably from the true values. At ratios of 15, \textsc{galfit} attempts to describe the source predominantly as a point source, yielding an unrealistically faint host magnitude ($\sim$32 mag). The estimated S\'ersic parameters hit the boundaries we set and exhibit extremely large uncertainties, indicating that these fits are unreliable.

Based on our decomposition results, all targets in our sample with detected host emission have AGN-to-host flux ratios below 2. In this regime, we expect \textsc{galfit} to recover a reliable S\'ersic profile for the host galaxy.

\subsection{The effects of host SNR}

Our fiducial galaxy GN1034159 has a SNR $>$ 10 in all bands, which is not the case for our full AGN sample. For the AGN galaxies in our sample with detected hosts, most have host S/N values above 3, with the exception of GN1086855 and GN1029154. In these two cases, although the host flux measured through aperture photometry has S/N $<$ 3, the S\'ersic model fluxes achieve S/N $\sim$ 3. We therefore test the host-recovery behaviour under lower SNR conditions. To reduce the host SNR, we increase the noise in both the science image and its error map. We then repeat the entire procedure described above: we first fit the original images to obtain a Sérsic model for the host galaxy, and then inject scaled AGN (point source) models to generate mock images with different AGN/host flux ratios.

We find that as the host SNR decreases, the recovery degrades more quickly as the AGN/host flux ratio increases. In other words, the upper limit of the AGN/host ratio for which we can still recover the host flux reliably becomes lower as the host SNR is reduced. In the SW bands, the recovery remains robust even when the SNR is reduced to $\sim$ 3. From F090W to F277W, the recovered host flux agrees with the true flux to within 20\%, with fractional uncertainties $\lesssim$ 40\%.

In contrast, the LW bands behave differently. In F356W, once the host SNR drops below $\sim$ 3, the recovered flux becomes meaningless: the fractional uncertainty exceeds 150\%, indicating that the measurement provides no useful constraint. In F444W, when the SNR falls below ~5, the recovered host flux becomes $\sim$50\% lower than the true value and the fractional uncertainty rises above $\sim$50\%; we therefore treat these measurements as unreliable.

Overall, we find that as SNR decreases and as AGN/host ratio increases, the recovery becomes progressively worse. However, for SNR $\gtrsim$ 3 and AGN/host ratios $\lesssim$ 1000, the host flux can still be reliably recovered in the F090W--F277W bands (bias $< 20\%$, fractional uncertainty $< 40\%$). For F356W and F444W, the useful information is lost when the host SNR drops below $\sim$5 and $\sim$3, respectively.

\begin{figure}[!ht]
\centering
    \includegraphics[width=\linewidth]{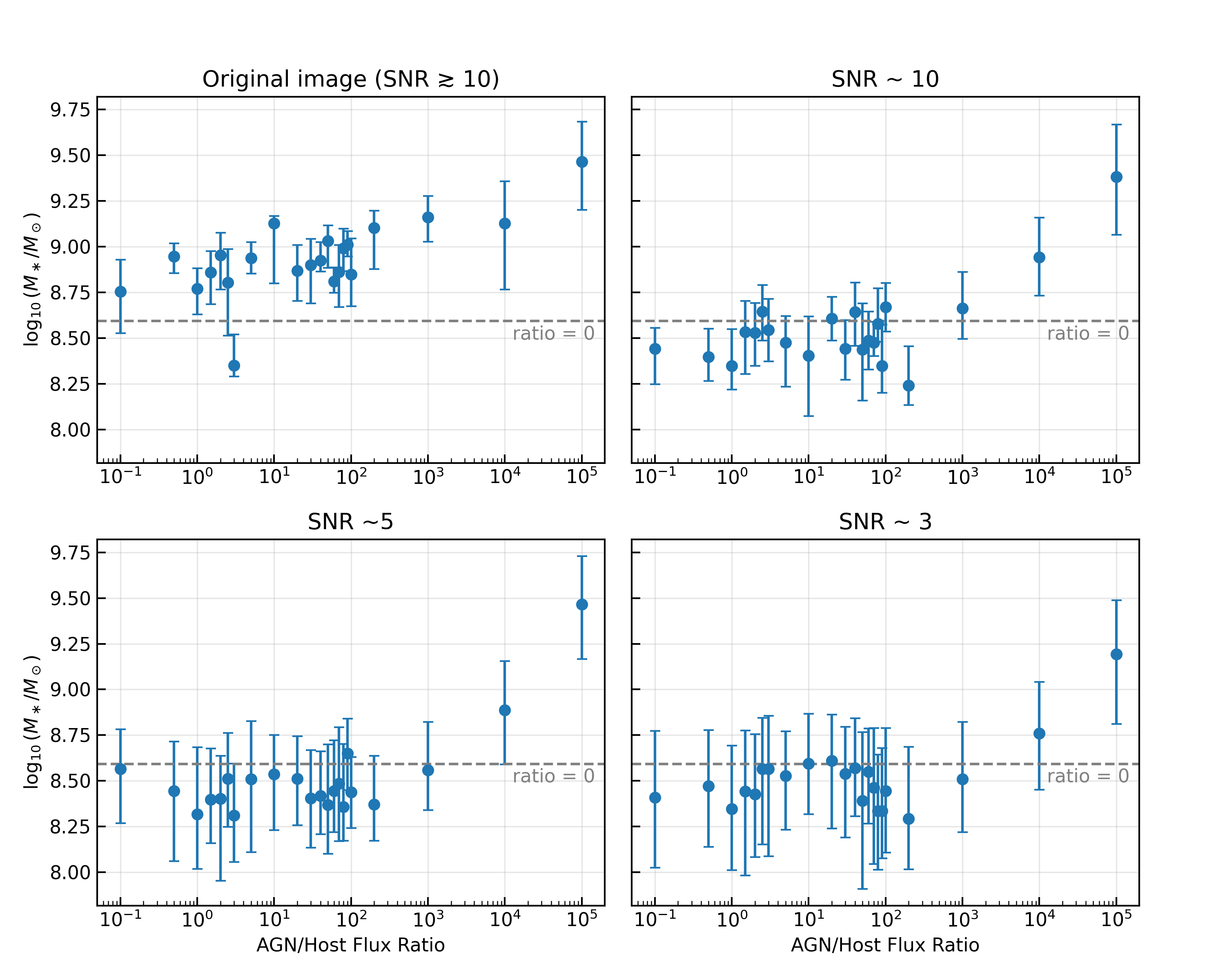}
\hfill
\caption{Stellar mass estimates as a function of the AGN-to-host flux ratio for simulated images at different SNRs of the host galaxy detection, with 1 $\sigma$ error bars. From top left to bottom right, the panels show results for the original image (SNR $\gtrsim$ 10), SNR $\sim$ 10, SNR $\sim$ 5, and SNR $\sim$ 3. The dashed line is the stellar mass estimate based on original input without AGN injection. Overall, the recovered stellar masses are broadly consistent with the non-injection case when AGN/host flux ratio is under 1000 and SNR is above 3. }
\label{fig:mass_recovery}
\end{figure}

What ultimately matters is whether the recovered fluxes allow us to obtain a reliable stellar-mass estimate. Running SED fits on the recovered fluxes for different SNR levels and AGN/host ratios, we find that as long as the AGN/host flux ratio is below $\sim$1000, the derived stellar mass differs from the true value (obtained from SED fits of the total flux of the original galaxy) by $\lesssim$ 0.5 dex, even when the host SNR is as low as $\sim$3, as shown in Figure~\ref{fig:mass_recovery}.

We therefore conclude that when the host SNR is above 3 and the AGN/host ratio is smaller than 1000, the stellar mass can be recovered within 0.5 dex. 

We also find that even when no AGN component is injected into the galaxy image, fitting the data with a PS+S\'ersic model would result in a recovered Sérsic flux systematically lower than the true total galaxy flux because some flux is always assigned to the PS component (the exact magnitude of the offset depends on both S/N and wavelength), corresponding to a typical stellar-mass uncertainty of $\approx$ 0.2 dex. This systematic offset is therefore intrinsic to the PS+S\'ersic decomposition, even in the absence of an AGN component.

\setcounter{figure}{0}
\renewcommand{\thefigure}{\thesection\arabic{figure}}

\clearpage

\bibliography{bib}{}
\bibliographystyle{aasjournalv7}

\end{document}